\documentclass[fleqn,usenatbib]{mnras}

\usepackage{newtxtext,newtxmath}

\usepackage[T1]{fontenc}

\DeclareRobustCommand{\VAN}[3]{#2}
\let\VANthebibliography\thebibliography
\def\thebibliography{\DeclareRobustCommand{\VAN}[3]{##3}\VANthebibliography}

\usepackage{graphicx}	
\usepackage{amsmath}	

\usepackage{diagbox}

\title[Mapping the Universe with Gamma-Ray Bursts]{Mapping the Universe with Gamma-Ray Bursts
}

\author[I. Horvath et al.]{
Istvan Horvath$^{1}$\thanks{E-mail: Istvan.Horvath@uni-nke.hu},
Zsolt Bagoly$^{1,2}$,
Lajos G. Balazs$^{3,4}$,
Jon Hakkila$^{5}$,
Zsuzsa Horvath$^{1}$
\newauthor 
Andras Peter Joo$^{4}$,
Sandor Pinter$^{1}$, 
L. Viktor T\'oth$^{4,6}$,
Peter Veres$^{7,8}$, 
and 
Istvan I. Racz$^{1}$
\\
$^{1}$University of Public Service, Budapest, Hungary\\
$^{2}$Department of Physics of Complex Systems, E\"otv\"os University, Budapest, Hungary \\
$^{3}$Konkoly Observatory, Research Centre for Astronomy and Earth Sciences, Budapest, Hungary\\
$^{4}$Department of Astronomy, E\"otv\"os Lor\'and University, Budapest, Hungary\\
$^{5}$Department of Physics and Astronomy, The University of Alabama in Huntsville, Huntsville, AL, USA  \\
$^{6}$University of Debrecen, Faculty of Science and Technology, Egyetem tér 1, H-4032 Debrecen, Hungary\\
$^{7}$Department of Space Science, University of Alabama in Huntsville, Huntsville, AL 35899, USA\\
$^{8}$Center for Space Plasma and Aeronomic Research, University of Alabama in Huntsville, Huntsville, AL 35899, USA
}

\date{Accepted XXX. Received YYc; in original form ZZZ}

\pubyear{2023}

\begin{document}
\label{firstpage}
\pagerange{\pageref{firstpage}-\pageref{lastpage}}
\maketitle

\begin{abstract}

We explore large-scale cosmic structure using the spatial distribution of 542 gamma-ray bursts (GRBs) having accurately-measured positions and spectroscopic redshifts. Prominent cosmological clusters are identified in both the northern and southern galactic hemispheres (avoiding extinction effects in the plane of the Milky Way) using the Bootstrap Point-Radius method. 
The Northern Galactic hemisphere contains a significant group of four GRBs in the redshift range $0.59 \le z \le 0.62$ (with a Bootstrap probability of $p=0.012$) along with the previously-identified Hercules--Corona Borealis Great Wall (in the revised redshift range $0.9 \le z \le 2.1$; $p=0.017$). The Southern Galactic hemisphere contains the previously-identified Giant GRB Ring ($p=0.022$) along with another possible cluster of $7-9$ GRBs at $1.17 \le z \le 1.444$ ($p=0.031$). 
Additionally, both the Hercules--Corona Borealis Great Wall and the Giant GRB Ring have become more prominent as the GRB sample size has grown.
The approach used here underscores the potential value of GRB clustering as a probe of large-scale cosmic structure, complementary to galaxy and quasar clustering. Because of the vast scale on which GRB clustering provides valuable insights, it is important that optical GRB monitoring continue so that additional spectroscopic redshift measurements should be obtained.

\end{abstract}




\section{Introduction}

Since their discovery, gamma-ray bursts (GRBs) have continued to pique the interest of scientists and captivate them with their enigmatic origins and remarkable energy outputs \citep{1973ApJ...182L..85K}. Despite the passage of decades, formulating a comprehensive theory that elucidates their genesis and satisfactorily explains every observed aspect remains an ongoing challenge \citep{2015AdAst2015E..22P,zhang_2018,2022GalaxBosPeer}. 

GRBs are widely acknowledged as one of the most intense and brilliantly luminous phenomena in the cosmos. They are believed to result from either the implosion of massive stars \citep{woo93,wb06} or from the merging of binary compact objects \citep{berger14}. Because of their extreme luminosities, a thorough examination of their
celestial coordinates, along with their intrinsic properties, holds promise to shed light on the Universe's largest-scale structures \citep{mp06,2019MNRAS.490.4481And,2020MNRAS.498.2544H,2023ApJ...952....3T}.

The GRBs' angular distribution was also
studied over the past few decades. 
Generally, GRBs have been found to be uniformly distributed on the sky 
\citep{Briggs96,bal98,bal99,mesz00,mgc03,vbh08,tarno15AA,tarno16MNRAS,2019ApJTarn,2019MNRAS.486.3027Ripa,2019MNRAS.490.4481And,2022AATarn},
although some subsamples appear to deviate from isotropy
\citep{bal98,Cline99,mbv00,mesz00,li01,mgc03,vbh08,R_pa_2017,2023MNRAS.524.3252H,2023CQGra..40i4001K}. Apart from the Hercules--Corona Borealis Great Wall identified by our group \citep{hhb14,hbht15}, 
one of the most significant recent developments in this field has been the detection of a large-scale structure in the redshift range of $0.78 < z < 0.86$, dubbed the "Giant GRB Ring", which appears to be  
consists of a circular arrangement of GRB positions at a distance of about 6 billion light-years \citep{BalazsRing2015}, although have been some debates about the existence of that structure \citep{2022AN....34320021F}. 
It appears to be somewhat smaller ($1.72$~Gpc) than the Hercules--Corona Borealis Great Wall ($2-3$~Gpc), but its presence in the data has been confirmed by further, elaborate statistical analysis \citep{BalazsTus2018}.
While the physical and astrophysical nature of both structures are not conclusively understood \citep[cf. the discussions in][]{BalazsRing2015,BalazsTus2018}, their existence may provide a challenge to standard assumptions about universal homogeneity and isotropy (i.e. the cosmological principle; see however \citealt{Li2015}). This means that studying structures like these is of high scientific importance \citep{2021A&A...653A.123M,2021MNRAS.507.1361M,2021A&A...649A.151M,2022SerAJ.204...29F}.

Aside from GRBs, other large structures have been discovered during the 21st century. These have included the Sloan Great Wall (which is a giant filament of galaxies) with a size of $\sim$\,0.4~Gpc \citep{Gott05}, the Huge Large Quasar Group with a size of 1.2~Gpc \citep{clo12}, and the Giant Quasar Arc with a size of 1~Gpc \citep{2022MNRASLopez}.
Both GRB-defined structures (the Giant GRB Ring and the Hercules--Corona Borealis Great Wall) exceed these in size.

Overall, these findings suggest that GRBs can serve as powerful probes of the large-scale structure of the Universe \citep{2023ApJ...952....3T}, providing valuable information about the distribution of matter and the evolution of cosmic structures over time \citep{2017AIPC.1792f0012R}.
Due to their large absolute brightness they can be
observed at very large cosmological distances. Consequently, the study of their locations in the sky
coupled with their properties can provide insights into the large-scale structure of the Universe.

\section{Data and Methods}

In this study we use a database of gamma-ray bursts (GRBs) having measured positions on the celestial sphere, optical afterglows, and spectroscopic redshifts; a majority of which were detected by NASA's Swift and Fermi experiments. Here we use the same dataset of \cite{2022Univ....8..221H} and \cite{universe8070342}, which is based on the dataset found in \cite{2020MNRAS.498.2544H}. The redshifts have been primarily obtained from the Gamma-Ray Burst Online Index (GRBOX) database; more than five hundred spectroscopic redshifts have been measured for GRBs as of today. It should be noted that, while the datasets were derived from different sources in the public domain, the GRBOX database heavily relies on notes from the Gamma-ray Coordination Network. GRBOX also relies on a publicly available dataset compiled by Joachim Greiner\footnote{https://www.mpe.mpg.de/~jcg/grbgen.html}, which provides extensive information on nearly all GRBs observed by any instrument. 
A thorough review was conducted during the assembly of the GRBOX catalog database\footnote{ Published by the Caltech Astronomy Department:  \url{http://www.astro.caltech.edu/grbox/grbox.php} } to exclude observations with significant systematic uncertainties. 
The redshift data used in \cite{2022Univ....8..221H,universe8070342} and \cite{2020MNRAS.498.2544H} exclusively consisted of spectroscopic redshifts, disregarding photometric redshifts and redshift estimates based on Ly-alpha limits because these data tend to introduce significant radial distance uncertainties. These uncertainties generally exceed several hundred Mpcs. 

We have supplemented the data from \cite{2022Univ....8..221H} and \cite{universe8070342} with new observations, obtained up until August 31st, 2022, that have resulted in a total of 542 GRBs with accurate position and spectroscopic redshifts (up to $z=8.26$).  We have used Joachim Greiner's table and relevant Gamma-ray Coordination Network messages to extend the dataset so that it includes the most recent data.

\subsection{The dataset}

The dataset used here is the most extensive one currently available for global studies, containing as it does precise GRB distances obtained from spectroscopic redshifts. It may be difficult to improve upon this data set, since the peak rate of GRBs with spectroscopic redshifts was $\approx$120/year after 2006, but has decreased rapidly since then, with each year yielding only $\approx 90\%$ of the previous year's redshift observations. This trend has only been disrupted in recent years by a few dedicated optical campaigns, but still, the current rate of new redshift observations remains embarrassingly low, preventing the possibility of significant statistical advancements in the near future \citep{universe8070342}.

\begin{figure}
    \centering
    \includegraphics[width=0.97\columnwidth]{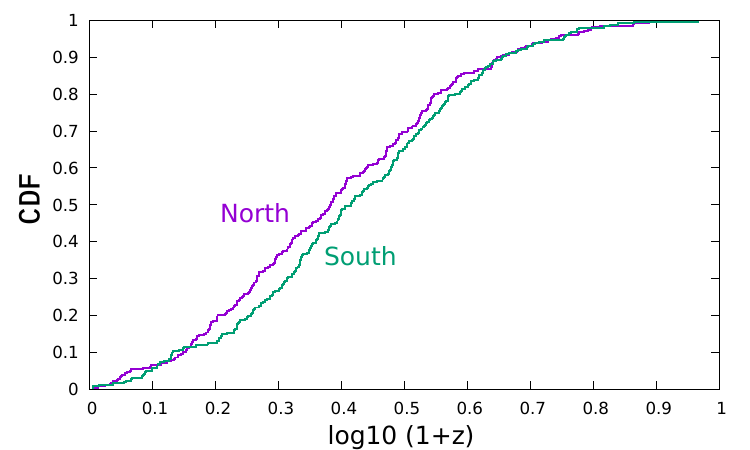}
    \caption{The cumulative distribution function (CDF) of log10(1+z) for GRBs in this study. Purple indicates the 262 GRBs found in the northern galactic sky, while green indicates the 280 GRBs found in the southern galactic sky. The difference between the two CDFs is insignificant (see text).}
    \label{fig:cumdist}
\end{figure}

\begin{figure}
    \centering
    \includegraphics[width=0.97\columnwidth]{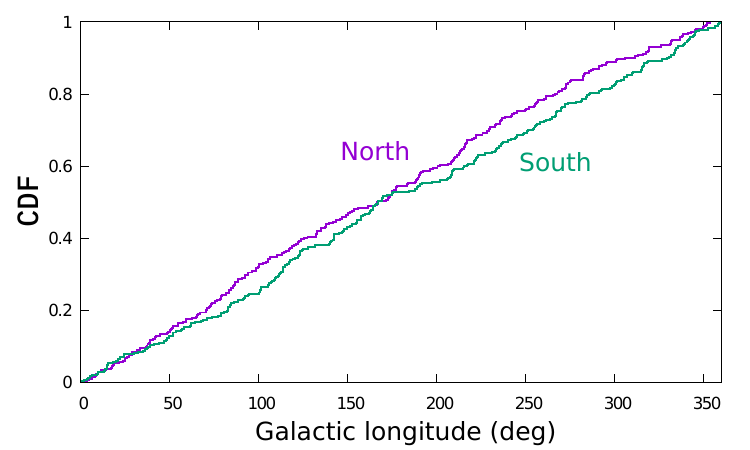}
    \caption{The cumulative distribution function (CDF) of the galactic longitudes for GRBs in this study. Purple (green) indicates the distributions of GRBs found in the northern (southern) galactic hemispheres.  The difference between the two CDFs is insignificant ($p=0.32$).}
    \label{fig:galacticl}
\end{figure}

Two GRBs have been removed from the dataset based on erroneous measurements. \cite{bagoly_horvath_hakkila_toth_2015} used the Spatial Two-Point Correlation Function to identify an interesting neighbouring pair of GRBs (GRB 020819B and GRB 050803). These two GRBs created a prominent peak in the Spatial Two-Point Correlation Function at a distance of $\approx 56$~Mpc. Interestingly, both GRBs' redshift values were later questioned in the literature (more than a decade after the events!). 
The Optically Unbiased GRB Host (TOUGH) survey measured a redshift of $z=3.5$ for the host galaxy of GRB 050803 which is completely inconsistent with the previous value of $z=0.422$ \citep{2015ApJ...808...73S}. Similarly, \cite{2017MNRAS.465L..89P} analyzed the spiral galaxy at a redshift of $z=0.41$ that was believed to be the host galaxy of GRB 020819B. By employing VLT MUSE and X-shooter observations, the authors concluded that the galaxy was an unrelated foreground galaxy. Consequently, given that both previous estimated distances are likely incorrect, we have removed these GRBs from the dataset. 

\begin{figure}
    \centering
    \includegraphics[width=0.97\columnwidth]{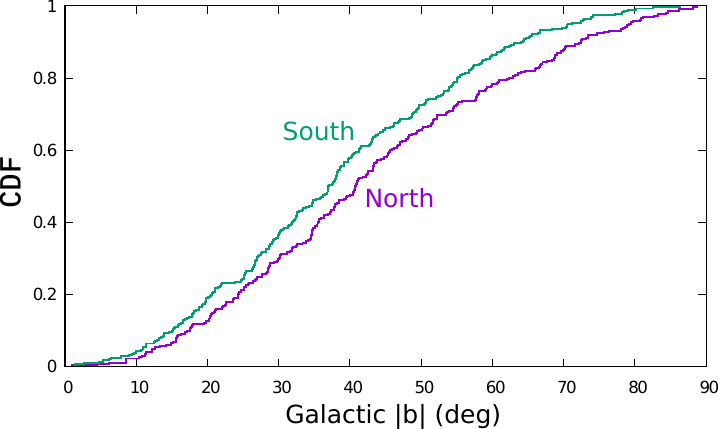}
    \caption{The cumulative distribution function (CDF) of the galactic latitude magnitudes for GRBs in this study. Purple (green) indicates the distributions of GRBs found in the northern (southern) galactic hemispheres.  The difference between the two CDFs is not significant.}
    \label{fig:galacticb}
\end{figure}

In the first three figures we demonstrate the distributions of spatial parameters. 
The cumulative distribution of the redshifts of the 542 GRBs can be seen in Fig.~\ref{fig:cumdist}. Of these, 262 are found in the northern galactic hemisphere, while the remaining 280 are found in the southern galactic hemisphere. The two-sample Kolmogorov-Smirnov test of these redshift distributions shows no significant difference between the two hemispheres ($D=0.098$, $p=0.151$). 
Fig.~\ref{fig:galacticl} shows the galactic longitude ($l$) coordinate distributions for both hemispheres. Because the maximum deviation of these distributions is $D=0.082$, the two-sample Kolmogorov-Smirnov test gives us a p-value of $p=0.325$, indicating that we cannot reject the hypothesis that the two samples are drawn from the same distribution. 
For the galactic longitude ($b$) coordinates, the KS test result similarly shows the two hemisphere distributions are not significantly different ($D=0.109$, $p=0.082$), see Fig.~\ref{fig:galacticb}.

\begin{figure}
    \centering
    \includegraphics[width=0.97\columnwidth]{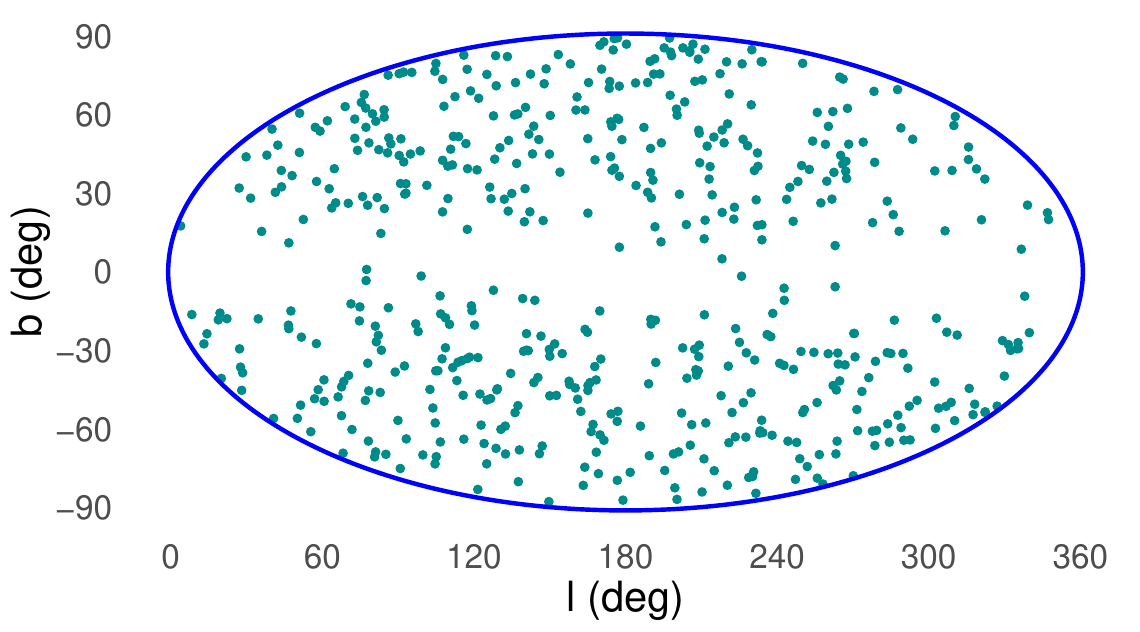}
    \caption{The sky distribution of the 542 analysed GRBs.}
    \label{fig:egen542}
\end{figure}

The sky distribution of the 542 GRBs can be seen in Fig.~\ref{fig:egen542}. There are only a very few spectroscopic redshift observations near the galactic plane as the visible gap for the low galactic $|b|$ coordinates shows. This reduced observational probability area will generate low bootstrap point-radius counts hence it will introduce large Poisson noise into the statistics. Avoiding this we analyze the two galactic hemispheres separately. This separation will reduce the method's sensitivity for any clustering around the galactic equator, where the low observational probability (which showed a low number of GRBs)
would anyway prevent any statistically reliable detection.

\subsection{Method of analysis}
\label{sec:method}
In order to test for the existence of clustering in the current database while
accounting for known sampling biases, we apply the bootstrap point-radius
method.  This method was described in Sec.~5. of \citet{hhb14}. A similar
analysis is performed here, albeit for a much larger dataset (as explained in
Sect.~\ref{sec:result}).
 
The bootstrap point-radius method searches for an overdensity in the point 
distribution.  We choose a slice of $n$ consecutive GRBs in their radial ($z$)
distribution, starting with position $k$, and determine the number of GRBs
within a volume defined by a spherical surface $A$. We refer to this surface as a {\em spherical cap}.  We look for the
$K$ maximum number of GRBs within the spherical cap, for all $k$ radial
starting position and for all spherical cap's center position on the sky,
therefore $K$ will depend on the given $A$ cap area and $n$ slice size only.

During the analysis we assume that the sky exposure is independent of~$z$ (e.g. 
this was shown in \cite{universe8070342}). Hence the same sky exposure factor was
assumed within a given spherical cap for the slice and the other GRBs outside
the slice, and differences in the spatial distribution could be analyzed.

For the spherical cap centers a Healpix partition \citep{2005ApJ...622..759G}
was chosen with $N=6$, providing 49152 quasi-isotropic positions.  The average
distance between them is $\approx 10$ times smaller than the average distance
between the GRBs, providing good coverage.

\begin{figure}
    \centering
    \includegraphics[width=0.9\columnwidth]{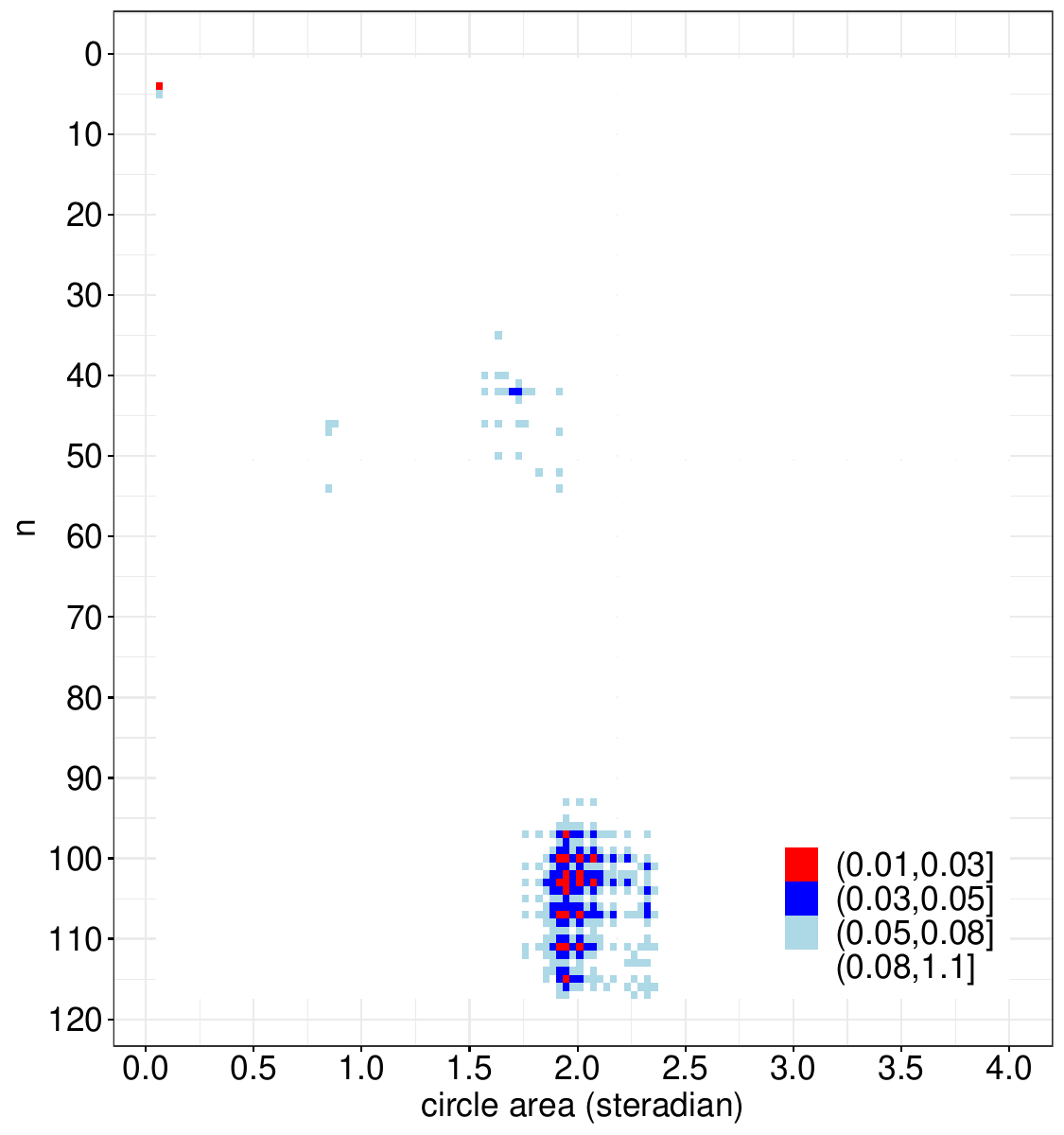}
    \caption{The bootstrap probabilities from the northern hemisphere on the (n, A) plane. White $p \ge 0.08$, light blue $0.08>p \ge 0.05$, blue $0.05>p \ge 0.03$, red $0.03>p \ge 0.01$. Only blue and red blocks are significant.
    }
    \label{fig:EszakProb}
\end{figure}

We do a Monte Carlo simulation by mixing the radial and angular positions 1000
times and repeating the process. Here we also select the largest number of GRBs
found ($K$) within a given spherical cap size $A$ and slice size $n$, and
compare them with the real $K$ value (for more details about this method see
our previous works, \citealt{hhb14,hbht15}, where the same method was applied).
We should mention that for similar $A$ and $n$ values the distribution is expected
to be similar (c.f.  Figs.~\ref{fig:EszakProb}. and \ref{fig:delprob}.).
  
The method will be sensitive to a density enhancement in a radial slice within
a given angular area. Clearly, the distribution of $K$ depends on the angular
two-point correlation function filtered by the spherical cap scale. The numbers
inside and outside the slice will depend non-linearly on the correlation
function because of the maximum cutoff. Also, the geometrical change in the angular
diameter distance with the cosmological co-moving distance makes the real
connection with the spatial two-point correlation function more complicated.

\section{Results}
\label{sec:result}

\subsection{Northern galactic hemisphere}
\label{sec:nhs}

Part of the results, the $K(n,A)$ function, for the northern galactic hemisphere can be seen in Table \ref{tab:n1}.
The complete $K(n,A)$ function can be found in the supplement materials. 
To calculate the probability of getting the number $K$ (for a certain $n$ and $A$), we apply the bootstrap method. We do this by retaining the sky position of the GRB and replacing its redshift with that of another GRB selected randomly from the sample. We do this for all 262 GRBs in the northern galactic hemisphere. We repeat the process as discussed in Sect.~\ref{sec:method} with this configuration and recalculate $K$ (with the same $n$ and $A$). We repeat this process 1000 times, counting how many times $K$ is greater than or equal to the measured value of $K$. We denote this number as $P$, and $P/1000$ estimates the probability of measuring a count of $K$ or larger.

Table \ref{tab:nprob} summarizes the probabilities of our analysis. Figure \ref{fig:EszakProb} shows also these results. 
There are three separated 
area in the $n$, $A$ plane
on which clustering is found to occur, indicative of significant deviations from isotropy/homogeneity. Notice that the light blue dots are not significant.

\begin{figure}
    \centering
    \includegraphics[width=0.95\columnwidth]{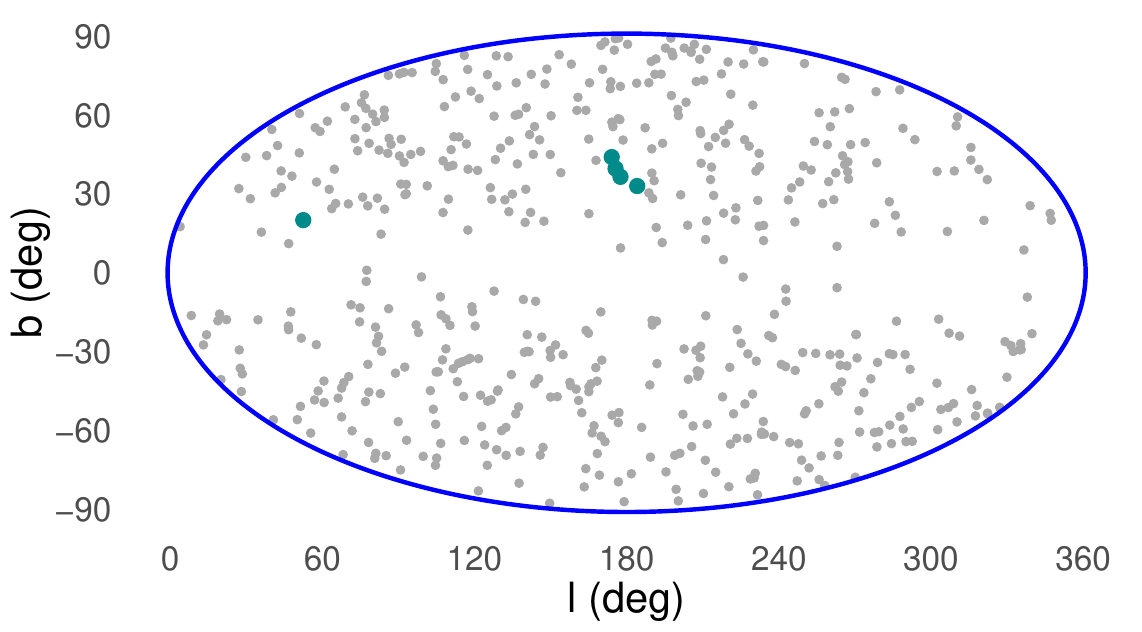}
    \caption{The 5 GRBs on the northern hemisphere within $0.59 \le z \le 0.62$.
    }
    \label{fig:4envonalban}
\end{figure}

The first of these 
occurs for $n=5$ and $A=0.0628$ ($0.02\times\pi$ steradian), with an observed $K=4$. Only one case ($5$, $0.0628$) reached the significant level in our analysis, with the bursts found within a redshift range of $0.59 \le z \le 0.62$. Fig. \ref{fig:4envonalban} shows the locations of these 5 GRBs on the sphere of the sky. We ran another 3000 bootstraps and only 47 runs (from the total 4000 runs) reached $K=4$, indicating that the probability of getting this fluctuation by chance is only $0.01175$. We also checked the $n=5$ and $A=0.0628$ parameters on the southern hemisphere. For 4000 bootstrap runs only 37 reached $K=4$, therefore the probability of getting this result by chance is even smaller ($p=0.00925$).
Table \ref{tab:5grb} shows the properties of these 5 GRBs.
Four out of the five GRBs in this group are positioned close to each other, raising the question of whether these GRBs or their host galaxies have something in common. We examined this by looking for similar properties of the GRBs and their host galaxies in literature data. We found $T90$ durations ranging from $\sim35-365 s$ classifying all these as long gamma-ray bursts (LGRBs), with peak energies of similar magnitudes $E_{p,obs}$ from $\sim95-137 keV$ (\cite{Xue_2019}, \cite{Tarnopolski_2021}). \cite{Bi_2018} calculated a star formation rate $SFR=81 yr^-1$ and a metallicity $log(Z/Z_{\odot})=-0.4$ for the host galaxy of GRB070612, supporting the high-mass collapsar model for LGRBs. GRB Coordinates Network reports, however, suggest differences in the metallicities of the four GRBs, with some metal lines only appearing in some of the spectra (Mg I, Mg II and Fe II in the case of GRB150323A, and Ca H and K, and C I in the case of GRB110106B (\cite{2011GCN.11538....1C}, \cite{2010GCN.10422....1C}, \cite{2007GCN..6556....1C}, \cite{2015GCN.17616....1P})).

\begin{table}
    \centering
    \begin{tabular}{l|l|c|r|r} 
    GRB name & z & d (Mpc) & l & b  \\  \hline
GRB150323A & 	0.593 & 	2209 & 174.802 & 	36.297        \\ \hline
GRB100213B & 	0.604 & 	2244 & 176.928 & 	33.339       \\ \hline
GRB050525A & 	0.6063 & 	 2251 & 54.953 & 	15.545        \\ \hline
GRB070612 & 	0.617 & 	 2285 & 183.626 & 	30.051        \\ \hline
GRB110106B & 	0.618 & 	2288 & 172.908 & 	40.449       \\ 

    \end{tabular}
    \caption{This table shows the 5 GRBs on the northern hemisphere in the $0.59 \le z \le 0.62$ redshift range. GRB ID, redshift (z), co-moving distance (d), galactic coordinates (l and b).}
    \label{tab:5grb}
\end{table}

The second 
suspicious area
is around $41 \le n \le 47$, $1.6 \le A \le 1.9$, with K being typically $29 \le K \le 32$. 
Although the bootstrap probabilities are close to be significant in a bigger (n, A) area ($36 \le n \le 57$, $1.57 \le A \le 1.95$), but only two neighbouring points reach the minimal significant limit (5\%); $n=43$, $A=1.7$, $p=0.038$ and $n=43$, $A=1.728$, $p=0.048$. 
This cluster is found in the redshift range $0.9 \le z \le 1.3$. Because it is 
only marginally significant ($p \le 0.04-0.08$), with only two points having 
small significance, we consider this cluster as only rising to the level of being suspicious. However, as we will see, the third area 
includes this part of the analysed parameter space in both redshift and sky position. 
It is worth noting that the sky area of this cluster corresponds to the position of the Hercules--Corona Borealis Great Wall \citep{hhb14,hbht15,2020MNRAS.498.2544H}.

The third area 
indicates anisotropies on a large angular scale, about $n=97-113$ and $1.9 \le A \le 2.1$, where the probabilities are less than $1-5\%$.  Table \ref{tab:Thw} shows the four most significant points in the ($n$, $A$) parameter space. 
For the $n=104$, $A=1.948$ we run another 3000 bootstraps and only 71 runs (from the total 4000 runs) reached $K=69$, indicating that the probability of getting this fluctuation by chance is only 0.01775.
Nineteen checked points had $p$ less than $3\%$ (red squares in Figure \ref{fig:EszakProb}).

\begin{figure}
    \centering
    \includegraphics[width=0.95\columnwidth]{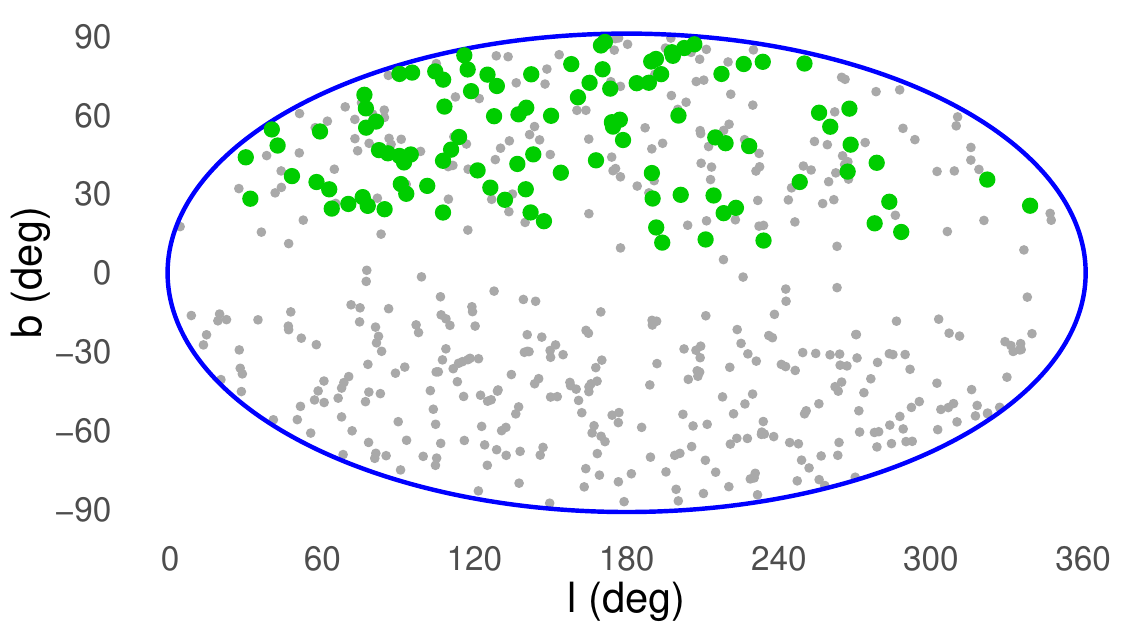}
    \caption{The northern hemisphere's (extended) Hercules-Corona Borealis Great Wall. The redshift range is $0.938 \le z \le 2.101$.
    }
    \label{fig:104agombon}
\end{figure}

Fig \ref{fig:104agombon} shows the 104 GRBs distribution on celestial sphere. Sixty nine of them are concentrated in $31\%$ of the hemisphere area.
The redshift range of this cluster is about $0.9 \le z \le 2.1$. This volume contains the previously-identified Hercules–Corona Borealis Great Wall (HCBGW) along with the second suspicious region (around $n=43-47$, $A=1.6-1.7$, $0.9 \le z \le 1.3$), which was discussed in this section previously. 

\begin{table}
    \centering
    \begin{tabular}{c|c|l|c|c|c} 
    n & A & p & K & z1 & z2 \\  \hline
101 & 	1.948 & 	0.02 & 67 & 	0.959	&	2.1      \\ \hline
104 & 	1.948 & \textcolor{red}{0.017} & 69 & 	0.938	&	2.1       \\ \hline
104 & 	2.011 & 	0.021 & 70 & 	0.938	&	2.1       \\ \hline
108 & 	1.948 & 	0.021 & 71 & 	0.885	&	2.1        \\ 

    \end{tabular}
    \caption{This table shows the 4 highest probabilities on the zone of $n=97-113$ and $A=1.9-2.1$. $n$ is the number of GRBs in the redshift range, $A$ is the circle size in steradian, $p$ is the bootstrap frequency, $K$ is the maximal GRB’s number in the circle, and $z1$ and $z2$ are the minimal and maximal redshifts.}
    \label{tab:Thw}
\end{table}

\subsection{Southern galactic hemisphere  }
\label{sec:shs}

\begin{figure}
    \centering
    \includegraphics[width=0.9\columnwidth]{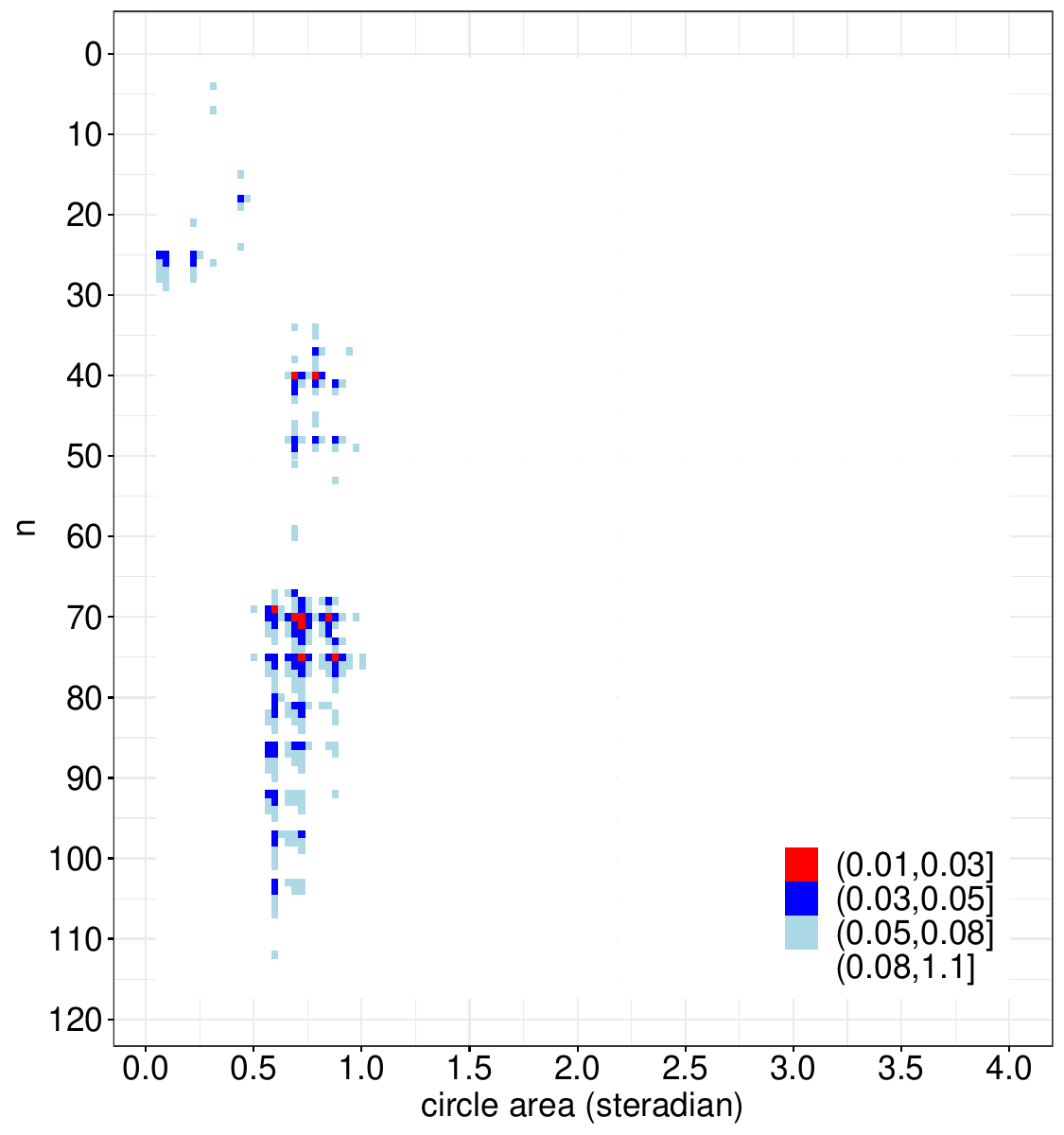}
    \caption{The bootstrap probabilities of the southern hemisphere on the ($n$, $A$) plane. White $p \ge 0.08$, light blue $0.08>p \ge 0.05$, blue $0.05>p \ge 0.03$, red $0.03>p \ge 0.01$. Only blue and red dots are significant.}
    \label{fig:delprob}
\end{figure}

The results for the southern galactic hemisphere can be
seen in Table \ref{tab:s1}, namely the $K$s.
The entire results of the analysis can be found in the supplement materials. 
The southern galactic hemisphere probabilities are shown in Table \ref{tab:sprob}.
Fig. \ref{fig:delprob} also demonstrates the bootstrap probabilities.

\begin{figure}
    \centering
    \includegraphics[width=0.95\columnwidth]{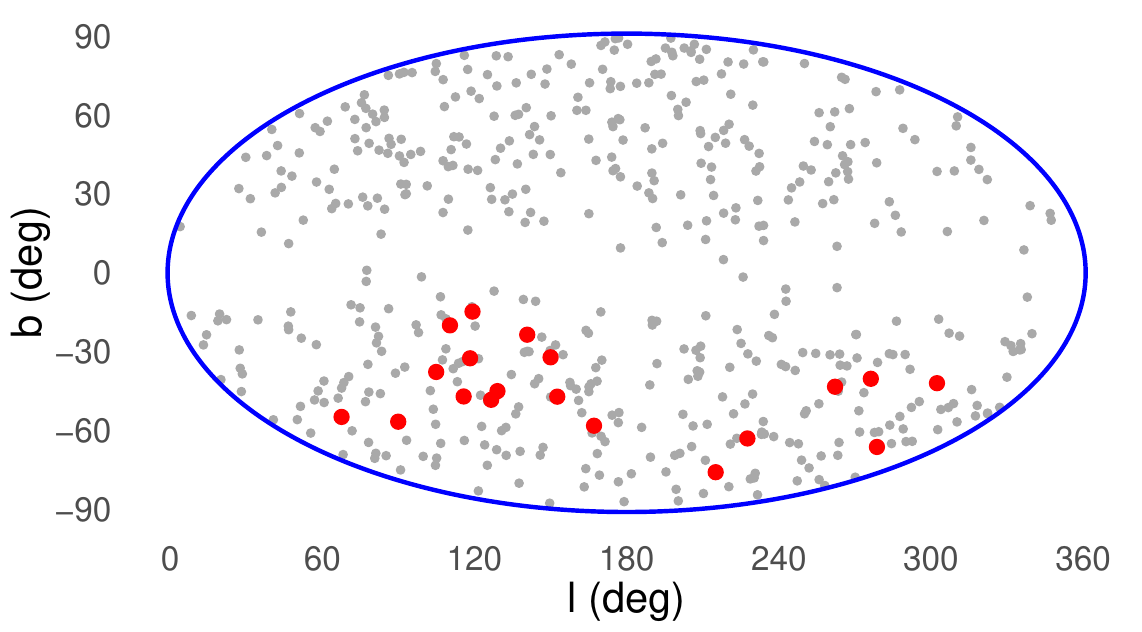}
    \caption{The figure shows the 19 GRBs in the $0.748-0.8595$ redshift range (red dots) in the southern hemisphere. This grouping was previously identified as the Giant GRB Ring.}
    \label{fig:ring}
\end{figure}

The first statistically interesting clustering is identified by $n=19$, $A=0.4396$, 
and $K=10$. The redshift range for this group is $0.748-0.8595$. This structure has been already discovered by \citet{BalazsRing2015}, and is shown in Fig. \ref{fig:ring}. 
In \citet{BalazsRing2015} the redshift range was $0.78 < z < 0.86$. Nine GRBs out of the 15 were in a small area. Now we have 10 GRBs out of the 19 in the same area.
The significance is $96.3\%$, see the blue dot in Fig. \ref{fig:delprob}, and the blue number in  Table \ref{tab:s1} (0.037 in $n=19$, $A=0.4396$). It is still significant, however not that much as \citet{BalazsRing2015} found.

\begin{figure} 
   \centering
  \includegraphics[width=0.95\columnwidth]{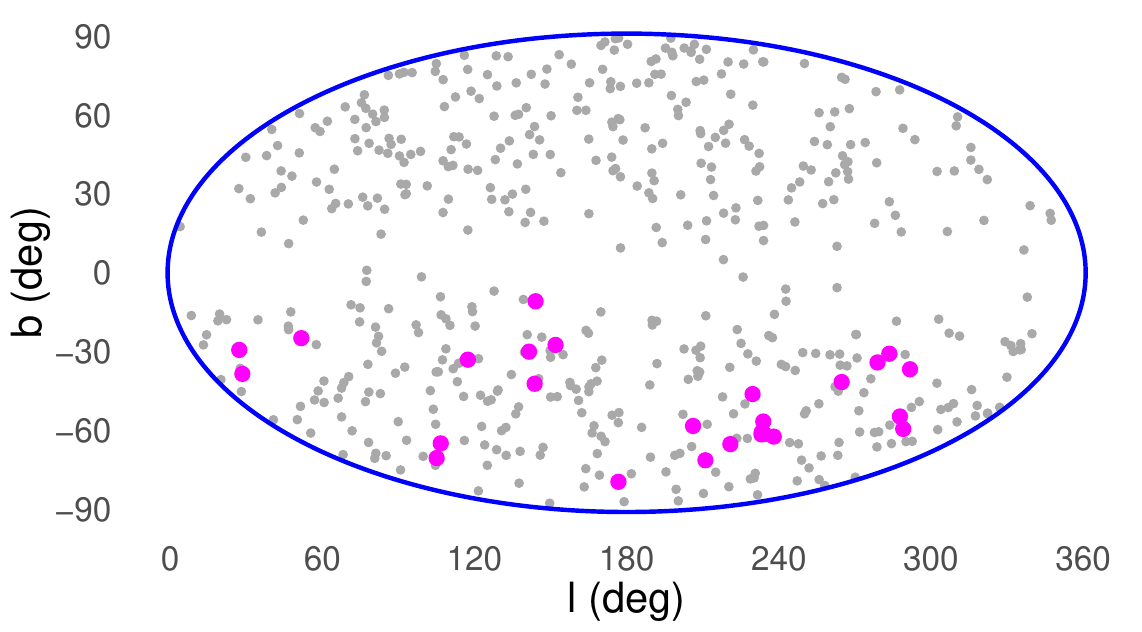}
  \caption{26 GRBs in the southern hemisphere, within $1.171 \le z \le 1.444$.}
   \label{fig:6GRB}
\end{figure}

The second statistically interesting southern hemisphere structure is found for $n=26$, $A=0.2199$, and $K=9$. The redshift range is $1.171 < z < 1.444$ and the significance is 96.9\%.
The second most significant value in this area is $p=0.035$ for $n=26$, $A=0.0942$, and $K=7$. The redshift range is the same. Notice that $n=26$ in both cases.
The sky distribution of this redshift range is shown in Fig. \ref{fig:6GRB}.
There is no previous reference to this cluster in the literature. However, due to the marginal significance and the small number of concentration (7 and 9) this could be only a statistical fluctuation.

\begin{figure} 
   \centering
  \includegraphics[width=0.95\columnwidth]{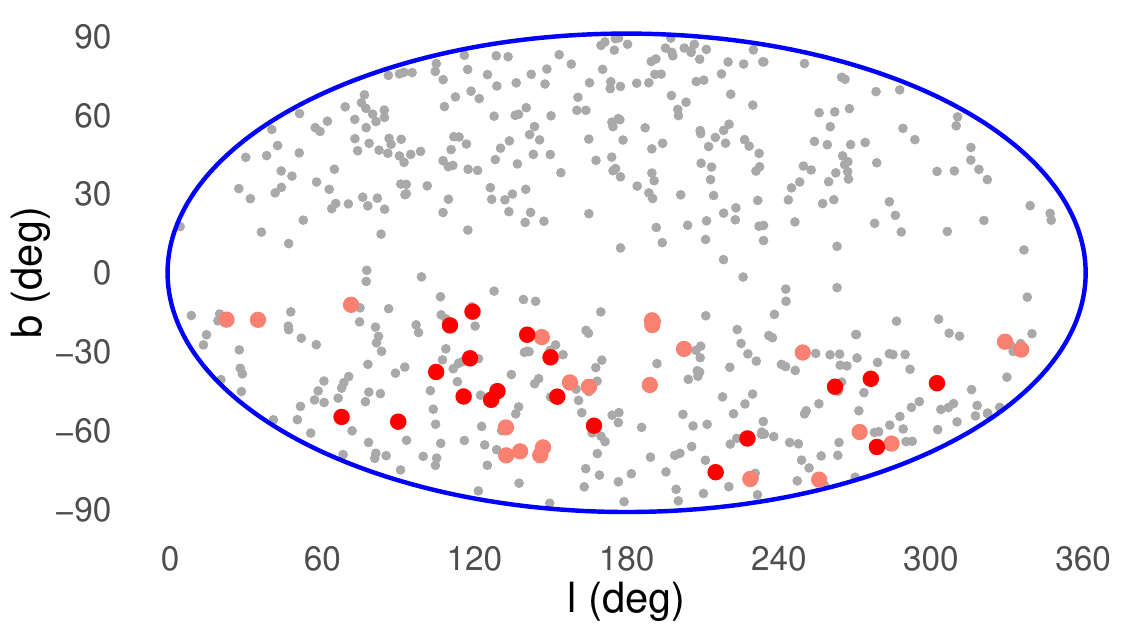}

 \caption{The southern hemisphere GRB sky distribution within $0.553 \le z \le 0.8595$, c.f. Fig.~\ref{fig:ring}. 
 The red dots are the same as in Fig.~\ref{fig:ring}, the salmon dots are within $0.553 \le z \le 0.748$. 
 A grouping around the Giant GRB Ring is growing. 
  }
   \label{fig:P553z8595}
\end{figure}

The next significant area is about $n=38-49$ and $A=0.69-0.88$. 
The two most significant part (the two red dots in Fig. \ref{fig:delprob}) are $n=41$, $A=0.691$, $K=17$, $p=0.024$ and $n=41$, $A=0.785$, $K=19$, $p=0.026$. In both cases the redshift range is $0.553 < z < 0.8595$ (see Fig. \ref{fig:P553z8595}).
As one can see this area contains the first significant area
which was about $0.748-0.8595$ redshift 
(the Giant GRB Ring), see red and salmon dots in Fig. \ref{fig:P553z8595}.

\begin{figure}
    \centering
    \includegraphics[width=0.95\columnwidth]{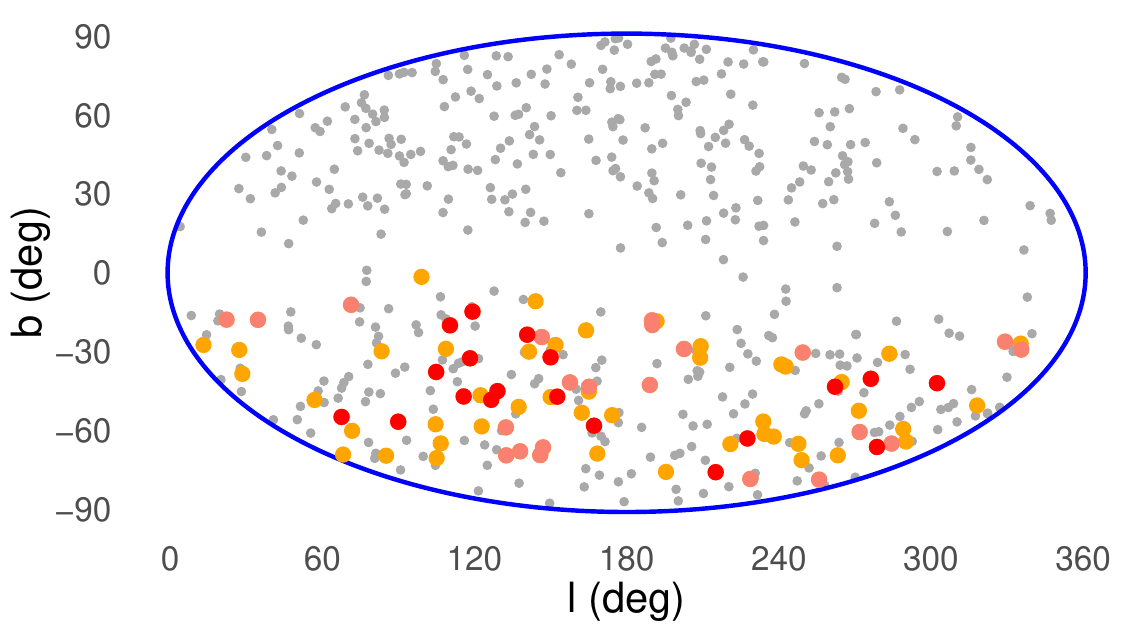}
    \caption{An even larger grouping around the Giant GRB Ring within  $0.553 < z < 1.35$. The area is about 0.7 steradian. C.f. Figs.~\ref{fig:ring}. and \ref{fig:P553z8595}.
    }
    \label{fig:GRB87}
\end{figure}

Finally, the southern hemisphere contains a large significant concentration of GRBs with n in the range $71 \le n \le 77$ and most of the concentration within an angular area of around 0.7 steradian.
The most significant points in this area of the $p(n,A)$ plane are listed in Table \ref{tab:nagydel}. Also see the 7 red dots in the left middle in Fig. \ref{fig:delprob}.
For the sky distribution of this cluster 
see Fig. \ref{fig:GRB87}

\begin{table}
    \centering
    \begin{tabular}{c|c|c|c|c|l} 
    n & A & K &  p & z1 & z2 \\  \hline
71 & 0.6908 & 24 & \textcolor{red}{0.022} &  0.553 & 1.155  \\ \hline
72 & 0.6908 & 24 & 0.027 &  0.553 & 1.170  \\ \hline
76 & 0.6908 & 25 & \textcolor{red}{0.023} &  0.553 & 1.233  \\ \hline
76 & 0.8792 & 29 & 0.025 &  0.553 & 1.233  \\ \hline
77 & 0.6908 & 25 & 0.027 &  0.553 & 1.25  \\ \hline
77 & 0.8792 & 29 & 0.031 &  0.553 & 1.25  \\ \hline
82 & 0.6908 & 26 & 0.027 &  0.553 & 1.31  \\ \hline
83 & 0.6908 & 26 & 0.035 &  0.553 & 1.325  \\ \hline
83 & 0.8792 & 30 & 0.047 &  0.553 & 1.325  \\ \hline
87 & 0.6908 & 27 & 0.031 &  0.553 & 1.350 \\ 

    \end{tabular}
    \caption{This table shows some points from the largest area of  
    significant field of the Figure \ref{fig:delprob}. $n$ is the number of GRBs in the redshift range, $A$ is the circle size in steradian, $K$ is the maximum number of GRBs in the circle, $p$ is the bootstrap frequency, and $z1$ and $z2$ are the minimal and maximal redshifts.}
    \label{tab:nagydel}
\end{table}

\section{North--South differences}

\begin{figure}
    \centering
    \includegraphics[width=\columnwidth]{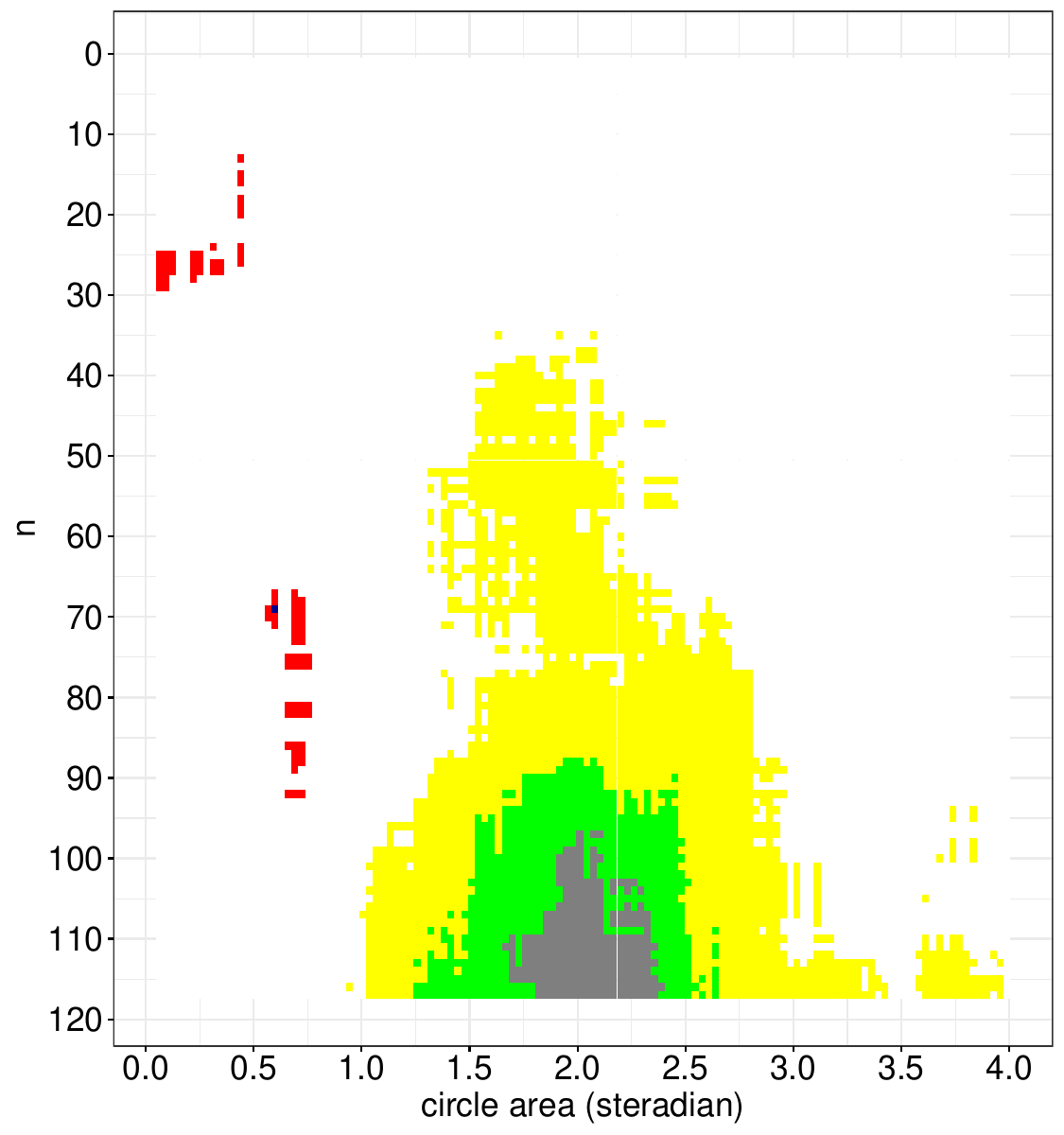}
    \caption{The difference between the $K_n$ northern and $K_s$ southern hemisphere $K(n,A)$ functions. 
    Blue $K_n-K_s = -3$, 
    red $K_n-K_s = -2$,
    white $-2 < K_n-K_s < 6$,
    yellow $5 < K_n-K_s < 11$,
    green $10 < K_n-K_s < 15$,
    grey $14 < K_n-K_s$.}
    \label{fig:KivonPboM}
\end{figure}

The isotropies of the two hemispheres can be compared to one another by subtracting Table \ref{tab:s1} from Table \ref{tab:n1}. The result, shown in Fig. \ref{fig:KivonPboM}, demonstrates that the two hemispheres exhibit similar, large-scale isotropy over most of the ($n$, $A$) plane, as indicated by the area colored white, where the differences are between -2 and 6. 
This means the observed K distributions in the ($n$, $A$) plane are similar in both hemispheres. 
However, there is a large region in which the difference between the distributions is three times larger than anywhere else. This suggests that, in the northern hemisphere, there is a large area in which there are significantly more GRBs than one would expect for a homogeneous, isotropic distribution.

\begin{figure}
    \centering
    \includegraphics[width=\columnwidth]{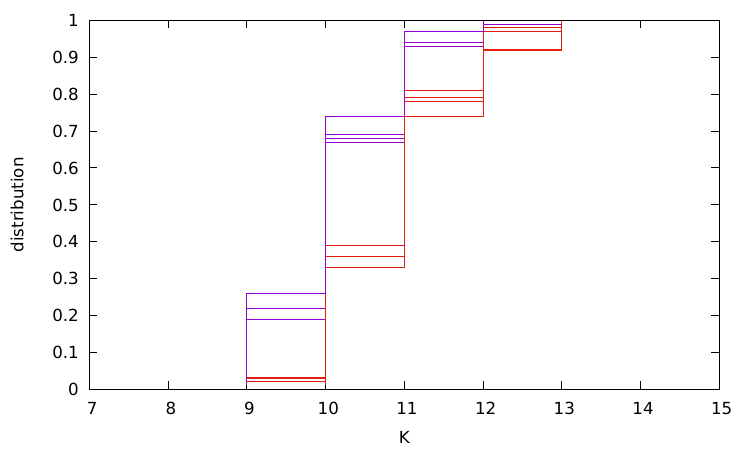}
    \caption{The difference between the northern (reds) and southern (blue) hemisphere K distributions in case $n=27$ and $A=0.503$. Each step function represents a hundred bootstrap runs. }
    \label{fig:n27A}
\end{figure}

The bootstrap probabilities can also be studied.
For example, in the case of $n=27$ and $A=0.503$, the number $K$ is 12 for both hemispheres.
The bootstrap probabilities are very similar. One can make 100 bootstrap runs and study the $K$ distribution. Every bootstrap run has a largest number, which we call $K$ (see Sect.~\ref{sec:method}). This 100 $K$s distribution can be seen in Figure \ref{fig:n27A}. In this plot, blue refers to northern hemisphere data while red refers to southern hemisphere data. In both cases, we plot 4 distributions (each having 100 runs). 
These runs do not differ very much from each other. This is also the case in the whole white area in Fig. \ref{fig:KivonPboM}. The K distributions are similar in most of the white area in the northern and southern hemispheres.
Over most of the sky, the differences between the northern and southern hemisphere $K$ distributions are minimal. However, that is not the case in the green and gray areas in Fig. \ref{fig:KivonPboM}. 

\begin{figure}
    \centering
    \includegraphics[width=\columnwidth]{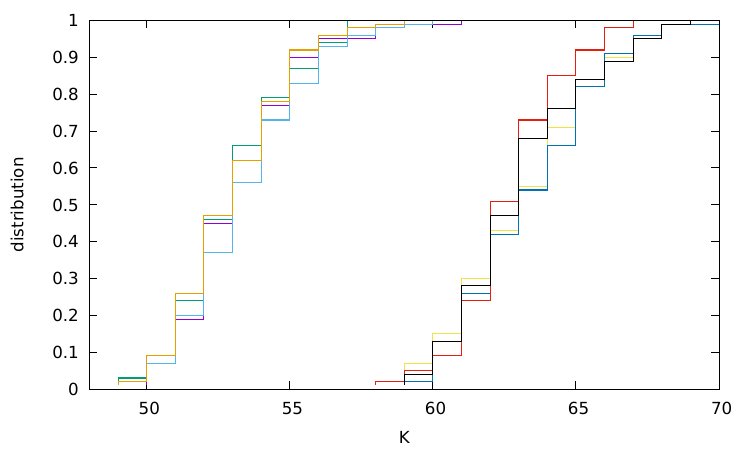}
    \caption{The difference between the northern (right) and southern (left) hemisphere K distributions in a case $n=104$ and $A=1.95$. }
    \label{fig:n104A}
\end{figure}

For example, the $K$ distributions are very different for $n=104$ and $A=1.95$, as seen in Figure \ref{fig:n104A}. 
In this case, for the observed 542 GRBs, $K$ is 69 in the northern hemisphere and 52 in the southern hemisphere.
The bootstrap frequency (reaching 69) in the northern hemisphere is 106 out of 5000 ($p=0.0212$). However, in the southern hemisphere, the bootstrap frequency $K$ never reaches 62, therefore the frequency for 69 is zero.
This demonstrates that the two hemisphere distributions differ very much around the 2-steradian scale, the northern hemisphere having about 17 more GRBs in this part of the sky.

\section{Discussion}

The discovery of the Sloan Great Wall (with a size of $\sim$\,0.4~Gpc) using galaxy locations \citep{Gott05} ushered in a new age of identifying large-scale universal structures via luminous objects.  The association of luminous quasars with galaxies has allowed even larger structures to be identified, including the Huge Large Quasar Group \citep{clo12} and the Giant Quasar Arc \citep{2022MNRASLopez} (each with a size of 1.2~Gpc).
GRBs, with their high luminosities, have extended this capability to even larger structures. Studying the spatial distribution of GRBs, both a great wall \citep{hhb14} and a Giant GRB Ring \citep{BalazsRing2015} were identified.

Because it is hard to collect unbiased survey data for galaxies and quasars, searches for large-scale groupings of these astronomical objects have relied on studies limited to specific spatial regions. 
However, since GRBs are mostly identified from large-scale surveys, it is easier to use them for examining large-scale deviations from isotropy and homogeneity over the full sky.
In this manuscript, we have shown how deviations from large-scale universal homogeneity can be studied using all available information 
pertaining to GRBs with known redshifts. 
Although the search for GRBs has taken place with great uniformity across the sky, a bias exists against measuring GRB redshifts close to the galactic plane.
Only a few GRBs with redshifts have been detectable in this region.
To minimize this known bias,
our approach is to separately study spatial distributions in
the northern and southern galactic hemispheres.

Using the method described in Sect. \ref{sec:method} we find two suspicious anisotropic regions in each hemisphere. 

\subsection{Northern hemisphere}

The northern hemisphere contains a cluster of five GRBs in the narrow redshift range ($0.59 \le z \le 0.62$) (see Table \ref{tab:5grb}).  Of these five, four are found in a small area (see Fig. \ref{fig:4envonalban}).

The second suspicious redshift range is about $0.9-1.3$, which contains $n=43-47$ GRBs, out of them $30-33$ is in an $A=1.6-1.7$ steradian size circle. These GRBs are in the same area of the sky, where the Hercules–Corona Borealis
Great Wall has been identified, except the HCBGW is farther away ($1.6<z<2.1$). However, the significance of this GRB group is weak in our present analysis.

The third grouping contains the second group ($0.9 < z < 1.3$) since it spans the $0.9-2.1$ redshift range. 
The sky position is the same for both case.
The bootstrap method shows significance in a huge area of the n, A parameter space. $97<n<113$ and $1.9<A<2.1$. For example, 69 out of the 104 GRBs are concentrated in the area around the Hercules and the Corona Borealis constellations in the sky.
The size of this volume (in co-moving distance) is $2.5\times3\times3$ Mpc or $8\times10\times10$ Gly, 
which contains 40\% more GRBs than expected. 

In conclusion, in the northern hemisphere we find a group of four GRBs and an extended space of the Hercules–Corona Borealis Great Wall. Based on data collected several years ago, the HCBGW appeared $3-4$ times smaller in radial size than we now measure it to be. Now we find the radial size to be comparable with the other two dimensions.

\subsection{Southern hemisphere}

There are four suspicious ranges in the southern hemisphere (see Fig. \ref{fig:delprob}). 

One is $n=26$, $A=0.2199$, and $K=9$. The redshift range is $1.171 < z < 1.444$ and the significance is $96.9\%$. See Fig. \ref{fig:6GRB} for this group which contains 9 GRBs. 
However, because of the small group number (9) and a weak significance ($p=0.031$) this could happened by chance. 

The other three significant ranges are boxed into each other.
Therefore these refer only to one GRB grouping.

As we discussed in Sect. \ref{sec:shs} the redshift ranges are
$0.748-0.8595$ ($K=10$),
$0.553 < z < 0.8595$ ($K=17$)
and 
$0.553 < z < 1.2-1.3$ ($K=24-29$).

One, which was previously found by \citep{BalazsRing2015}, with a redshift $0.75-0.86$, containing 10 GRBs out of 19 on 0.4396 steradian in the sky. The second contains 
17 GRBs (out of 41) including the previous 10. 
The third group contains $24-29$ GRBs out of $71-77$ on a redshift range from 0.55 to about $1.2-1.3$ covering 0.7 steradian (see Table \ref{tab:nagydel}).
This volume contains the previous two volumes.

The techniques described in this manuscript have identified probable large-scale anisotropic cosmological structures using luminous GRBs.
As the GRB sample size has increased, these structures have become more pronounced and new structures are beginning to appear. These results support the continued monitoring of the sky for GRBs using survey satellites.  

\subsection{Binomial probabilities and number of tries}

The Binomial Theorem can be used as an alternate approach to obtaining clustering probabilities. 
For data tested many times, the overall binomial probability is found by multiplying the individual binomial probability obtained for a measurement by the number of independent trials.
However, it is hard to directly use the number of bootstrap trials to calculate a binomial probability because a great many of these trials are not independent. We might avoid this problem by employing Monte-Carlo analysis and re-analyzing the data instead of identifying independent trials from our bootstrap procedure, but the latter approach is preferred because uncertainty in the sky exposure function makes the Monte Carlo approach a less reliable technique for this analysis.

The southern sky contains 280 GRBs with measured redshifts. For every circle size containing $n=100$ points, we made trials for 180 intervals. Since these intervals overlap, only a handful of them are independent. For example, the $k=1$ and $k=2$ intervals contain 100 GRBs of which 99 are the same. For $n=5$ there are 275 intervals, of which only 56 are totally independent. Similarly, we tried 116 different $n$ values ($3 \le n \le 118$), but these were not independent because every trial with $n=16$ included two trials with $n=15$ and three trials with $n=14$, etc. Since there are only 280 southern hemisphere GRBs in our sample, we can estimate the number of trials as being a few hundred.

Due to overlap, we must similarly estimate the number of independent spherical cap trials. We made 10,000 position trials representing every size of an angular circle. For a circle with area 0.1 steradian, some 50 could be categorized as being independent. For a circle of area 0.5 steradian, only 9-10 can be independent, and these are not independent from many previous trials like the 0.44 or the 0.56 steradian circle sizes. Therefore, we can estimate the number of trials also as a few hundred. Multiplying this with the number obtained in the previous paragraph (also a few hundred) the number of trials is estimated as being between $5\times 10^4$ and $2\times 10^5$. 

With this information we can estimate the binomial probabilities for our prominent clusters. In the northern hemisphere, the first significant result occurs for $n=5$, $A=0.0628$, and $K=4$. The binomial probability with 5, 4 and $p=0.01$ (because $A=0.0628$ is one hundredth of the half sky) is $5\times 10^{-8}$. Since we made $5\times 10^4$ or $2\times 10^5$ trials, the probability is still no larger than $p=0.01$.
For the case of $n=43$, $A=1.7$, and $K=30$, the individual $p=0.27$ and the binomial probability (for 43, 30) is $6.9\times 10^{-9}$. Multiplying this by $5\times 10^4$ or $2\times 10^5$ results in $p=(4-14) \times 10^{-4}.$
For the case of $n=104$, $K=69$, and $A=1.948$, we find $p=0.31$, for which the binomial probability is $10^{-14}$, which is still smaller than $10^{-8}$.
Table~\ref{tab:proN} contains the estimated binomial and bootstrap probabilities for each of these significant clusters. 

Table~\ref{tab:proS} contains the similarly-calculated binomial estimated probabilities for the southern hemisphere.
It can be seen the binomial probabilities are generally similar to or slightly less than than the bootstrap probabilities. 

The last four entries in Table~\ref{tab:proN} identify binomial probabilities which are much more smaller than the corresponding bootstrap probabilities. As the table and 
Fig.~\ref{fig:EszakProb} show, there are 87 more trials that show this very low estimated probabilities.

The Cosmological Principle implies that the redshift and sky distribution of GRBs in the northern and southern galactic hemispheres should be similar. However, Fig.~\ref{fig:KivonPboM} demonstrates that they are not. 
Although more southern galactic GRBs have been detected at present, many of the northern sky GRBs are found within the large cluster located at $1 \le z \le 2$. Bootstrapping events from the northern sky causes us to choose many GRBs from this concentration, which means most of the bootstrapped samples contain many GRBs from this group. Fig.~\ref{fig:n104A} demonstrates this effect, where the average bootstrap numbers are larger (in average) with 10 in the northern hemisphere than in the southern hemisphere. 
Other than this, Fig.~\ref{fig:n27A} demonstrates that most samples in the two hemispheres that are bootstrapped $K$ (with the same $n$ and $A$) do not differ by much.
This suggests that the bootstrap probabilities tend to overestimate the significance of features relative to the values that are obtained using the binomial estimate.

\begin{table}
    \centering
    \begin{tabular}{r|r|l|c|c|r|c} 
n & K & Arel & $p_{binom}$ & $p_{est}$ & num & $p_{bstr}$   \\  \hline
5 & 4 &  0.01 &  $ 5\times 10^{-8}$ &  0.0025-0.01 & 1 & 0.021  \\ \hline
43 & 30 &  0.27 &$ 7\times 10^{-9}$ &  0.0004-0.0014 & 2 & 0.038  \\ \hline
101 & 67 &  0.31 & $ 2\times 10^{-13}$ & $ (1-4)\times 10^{-8}$ & 91 & 0.02  \\ \hline
104 & 69 &  0.31 & $ 1\times 10^{-14}$ & < $10^{-8}$ & 91 & 0.017  \\ \hline
104 & 70 &  0.32 & $ 5\times 10^{-14}$ & <$10^{-8}$ & 91 & 0.021  \\ \hline
108 & 71 &  0.31 & $ 5\times 10^{-14}$ & <$10^{-8}$ & 91 & 0.021  \\ 

    \end{tabular}
    \caption{Binomial probabilities and the estimated probabilities (multiplied by the number of tries) for the northern hemisphere ($p_{est}$). K and n are the same as described in Sect.~\ref{sec:method}. Here, Arel is the relative area of the hemisphere,
    $p_{binom}$ is the binomial probability with p=Arel, n tries, and K successes. $p_{est}$ is the binomial probability multiplied by the number of independent tries, 'num' is the number of significant signs in the  
   neighboring area shown in Fig.~\ref{fig:EszakProb}, and 
    $p_{bstr}$ is the bootstrap probability calculated in Sect.~\ref{sec:nhs}. Some of these are shown in Table~\ref{tab:Thw}. 
    }
    \label{tab:proN}
\end{table}

\begin{table}
    \centering
    \begin{tabular}{r|r|l|c|c|r|c} 
n & K & Arel & $p_{binom}$  & $p_{est}$ & num & $p_{bstr}$   \\  \hline
19 & 10 &  0.07 & $ 1.45\times 10^{-7}$ & 0.007-0.03 & 1 & 0.037  \\ \hline
26 & 9 &  0.035 & $ 1.43\times 10^{-7}$ & 0.007-0.03 & 5 & 0.031  \\ \hline
26 & 7 &  0.015 & $ 9\times 10^{-8}$ & 0.005-0.018 & 5 & 0.035  \\ \hline
41 & 19 &  0.125 & $ 1.06\times 10^{-7}$ & 0.005-0.02 & 13 & 0.026  \\ \hline
71 & 24 &  0.11 & $ 2.8\times 10^{-7}$ & 0.014-0.05 & 60 & 0.022  \\ \hline
72 & 24 &  0.11 & $ 3.8\times 10^{-7}$ & 0.019-0.07 & 60 & 0.027  \\ \hline
76 & 25 &  0.11 & $ 2.9\times 10^{-7}$ & 0.015-0.06 & 60 & 0.023  \\ \hline
77 & 25 &  0.11 & $ 3.88\times 10^{-7}$ & 0.019-0.07 & 60 & 0.027  \\ \hline
82 & 26 &  0.11 & $ 3.87\times 10^{-7}$ & 0.019-0.07 & 60 & 0.027  \\ \hline
83 & 26 &  0.11 & $ 5.04\times 10^{-7}$ & 0.025-0.1 & 60 & 0.035  \\ \hline
87 & 27 &  0.11 & $ 3.79\times 10^{-7}$ & 0.019-0.07 & 60 & 0.031       \\ 

    \end{tabular}
    \caption{Binomial probabilities and the estimated probabilities (multiplied by the number of tries) for the southern hemisphere ($p_{est}$). K and n are the same as described in Sect.~\ref{sec:method}. Here, Arel is the relative area of the hemisphere,
    $p_{binom}$ is the binomial probability with p=Arel, n tries, and K successes. $p_{est}$ is the binomial probability multiplied by the number of tries, 'num' is the number of significant signs in the neighboring areas shown in Fig.~\ref{fig:delprob}.
    $p_{bstr}$ is the bootstrap probabilities which were calculated in Sect.~\ref{sec:shs}. Some of these are shown in Table~\ref{tab:nagydel}. }
    \label{tab:proS}
\end{table}

\section{Summary}

Our method investigated the 542 GRBs having observed redshift. We vary the redshift, the group size, and the area of the sky. In this 3D parameter space we found several suspicious regions. One in the northern galactic sky contains 4 GRBs with $1.2\%$ probability of having this by chance (see Fig. \ref{fig:4envonalban}). 
In the southern hemisphere there is another small group having 7 or 9 GRBs, but the significance in both cases does not reach even the $3\%$ level (see Fig. \ref{fig:6GRB}). 

However, there were two groups (one in each hemisphere), 
where the clustering yielded several significant results in the ($n,A$) plane (red blocks in Figures \ref{fig:EszakProb} and \ref{fig:delprob}).  
In the southern hemisphere there are three areas in the ($n$, $A$) space. From these areas one area's small part was previously identified as the Giant GRB Ring \citep{BalazsRing2015,BalazsTus2018} (see Figures \ref{fig:ring}, \ref{fig:P553z8595}, and \ref{fig:GRB87}). In the northern hemisphere a huge part of the ($n$, $A$) plane suggested a group (having $69-71$ GRBs out of the $104-108$, see Table \ref{tab:Thw}) in a $31\%$ area of the hemisphere (Fig. \ref{fig:104agombon}). A part of this structure was identified previously in the literature as the Hercules--Corona Borealis Great Wall \citep{hhb14,hbht15}.

\section*{Acknowledgements}

The authors thank the Hungarian TKP2021-NVA-16 and
OTKA K134257 program for their support. 
The authors also thank Dr Robert J. Nemiroff for useful comments that improved the paper.

\section*{Data Availability}

The data underlying this paper are available in Gamma-Ray Burst Online Index (GRBOX) database published by the Caltech Astronomy Department  (\url{http://www.astro.caltech.edu/grbox/grbox.php}),
Joachim Greiner's table (\url{https://www.mpe.mpg.de/~jcg/grbgen.html}), 
and relevant Gamma-ray Coordination Network (\url{https://gcn.gsfc.nasa.gov/gcn3_archive.html}) messages. The code and data used for this study are available at \url{https://github.com/zbagoly/GRBSphericalCapStat}.



\bibliographystyle{mnras}
\bibliography{main,lajos,horv22b}

\hyphenation{Post-Script Sprin-ger}
\begin{thebibliography}{}
\makeatletter
\relax
\def\mn@urlcharsother{\let\do\@makeother \do\$\do\&\do\#\do\^\do\_\do\%\do\~}
\def\mn@doi{\begingroup\mn@urlcharsother \@ifnextchar [ {\mn@doi@} {\mn@doi@[]}}
\def\mn@doi@[#1]#2{\def\@tempa{#1}\ifx\@tempa\@empty \href {http://dx.doi.org/#2} {doi:#2}\else \href {http://dx.doi.org/#2} {#1}\fi \endgroup}
\def\mn@eprint#1#2{\mn@eprint@#1:#2::\@nil}
\def\mn@eprint@arXiv#1{\href {http://arxiv.org/abs/#1} {{\tt arXiv:#1}}}
\def\mn@eprint@dblp#1{\href {http://dblp.uni-trier.de/rec/bibtex/#1.xml} {dblp:#1}}
\def\mn@eprint@#1:#2:#3:#4\@nil{\def\@tempa {#1}\def\@tempb {#2}\def\@tempc {#3}\ifx \@tempc \@empty \let \@tempc \@tempb \let \@tempb \@tempa \fi \ifx \@tempb \@empty \def\@tempb {arXiv}\fi \@ifundefined {mn@eprint@\@tempb}{\@tempb:\@tempc}{\expandafter \expandafter \csname mn@eprint@\@tempb\endcsname \expandafter{\@tempc}}}

\bibitem[\protect\citeauthoryear{{Andrade}, {Bengaly}, {Alcaniz}  \& {Capozziello}}{{Andrade} et~al.}{2019}]{2019MNRAS.490.4481And}
{Andrade} U.,  {Bengaly} C. A.~P.,  {Alcaniz} J.~S.,   {Capozziello} S.,  2019, \mn@doi [\mnras] {10.1093/mnras/stz2754}, \href {https://ui.adsabs.harvard.edu/abs/2019MNRAS.490.4481A} {490, 4481}

\bibitem[\protect\citeauthoryear{{Bagoly}, {Horv{\'a}th}, {Hakkila}  \& {T{\'o}th}}{{Bagoly} et~al.}{2015}]{bagoly_horvath_hakkila_toth_2015}
{Bagoly} Z.,  {Horv{\'a}th} I.,  {Hakkila} J.,   {T{\'o}th} L.~V.,  2015, \mn@doi [Proceedings of the International Astronomical Union] {10.1017/S1743921315010182}, 11, 2–2

\bibitem[\protect\citeauthoryear{Bagoly, Horvath, Racz, Balázs  \& Tóth}{Bagoly et~al.}{2022}]{universe8070342}
Bagoly Z.,  Horvath I.,  Racz I.~I.,  Balázs L.~G.,   Tóth L.~V.,  2022, \mn@doi [Universe] {10.3390/universe8070342}, 8, 342

\bibitem[\protect\citeauthoryear{{Bal{\'a}zs}, {M{\'e}sz{\'a}ros}  \& {Horv{\'a}th}}{{Bal{\'a}zs} et~al.}{1998}]{bal98}
{Bal{\'a}zs} L.~G.,  {M{\'e}sz{\'a}ros} A.,   {Horv{\'a}th} I.,  1998, \aap, \href {http://adsabs.harvard.edu/abs/1998A%26A...339....1B} {339, 1}

\bibitem[\protect\citeauthoryear{{Bal{\'a}zs}, {M{\'e}sz{\'a}ros}, {Horv{\'a}th}  \& {Vavrek}}{{Bal{\'a}zs} et~al.}{1999}]{bal99}
{Bal{\'a}zs} L.~G.,  {M{\'e}sz{\'a}ros} A.,  {Horv{\'a}th} I.,   {Vavrek} R.,  1999, \aaps, \href {http://adsabs.harvard.edu/abs/1999A%26AS..138..417B} {138, 417}

\bibitem[\protect\citeauthoryear{{Bal{\'a}zs}, {Bagoly}, {Hakkila}, {Horv{\'a}th}, {K{\'o}bori}, {R{\'a}cz}  \& {T{\'o}th}}{{Bal{\'a}zs} et~al.}{2015}]{BalazsRing2015}
{Bal{\'a}zs} L.~G.,  {Bagoly} Z.,  {Hakkila} J.~E.,  {Horv{\'a}th} I.,  {K{\'o}bori} J.,  {R{\'a}cz} I.~I.,   {T{\'o}th} L.~V.,  2015, \mn@doi [\mnras] {10.1093/mnras/stv1421}, \href {https://ui.adsabs.harvard.edu/abs/2015MNRAS.452.2236B} {452, 2236}

\bibitem[\protect\citeauthoryear{{Bal{\'a}zs}, {Rejt{\H{o}}}  \& {Tusn{\'a}dy}}{{Bal{\'a}zs} et~al.}{2018}]{BalazsTus2018}
{Bal{\'a}zs} L.~G.,  {Rejt{\H{o}}} L.,   {Tusn{\'a}dy} G.,  2018, \mn@doi [\mnras] {10.1093/mnras/stx2550}, \href {https://ui.adsabs.harvard.edu/abs/2018MNRAS.473.3169B} {473, 3169}

\bibitem[\protect\citeauthoryear{{Berger}}{{Berger}}{2014}]{berger14}
{Berger} E.,  2014, \mn@doi [\araa] {10.1146/annurev-astro-081913-035926}, \href {http://adsabs.harvard.edu/abs/2014ARA%26A..52...43B} {52, 43}

\bibitem[\protect\citeauthoryear{Bi, Mao, Liu  \& Bai}{Bi et~al.}{2018}]{Bi_2018}
Bi X.,  Mao J.,  Liu C.,   Bai J.-M.,  2018, \mn@doi [The Astrophysical Journal] {10.3847/1538-4357/aadcf8}, 866, 97

\bibitem[\protect\citeauthoryear{{Bo{\v{s}}njak}, {Barniol Duran}  \& {Pe'er}}{{Bo{\v{s}}njak} et~al.}{2022}]{2022GalaxBosPeer}
{Bo{\v{s}}njak} {\v{Z}}.,  {Barniol Duran} R.,   {Pe'er} A.,  2022, \mn@doi [Galaxies] {10.3390/galaxies10020038}, \href {https://ui.adsabs.harvard.edu/abs/2022Galax..10...38B} {10, 38}

\bibitem[\protect\citeauthoryear{{Briggs} et~al.,}{{Briggs} et~al.}{1996}]{Briggs96}
{Briggs} M.~S.,  et~al., 1996, \mn@doi [\apj] {10.1086/176867}, \href {http://adsabs.harvard.edu/abs/1996ApJ...459...40B} {459, 40}

\bibitem[\protect\citeauthoryear{{Cenko}, {Fox}, {Cucchiara}, {Schmidt}, {Berger}, {Price}  \& {Roth}}{{Cenko} et~al.}{2007}]{2007GCN..6556....1C}
{Cenko} S.~B.,  {Fox} D.~B.,  {Cucchiara} A.,  {Schmidt} B.~P.,  {Berger} E.,  {Price} P.~A.,   {Roth} K.~C.,  2007, GRB Coordinates Network, \href {https://ui.adsabs.harvard.edu/abs/2007GCN..6556....1C} {6556, 1}

\bibitem[\protect\citeauthoryear{{Chornock}, {Berger}  \& {Fox}}{{Chornock} et~al.}{2011}]{2011GCN.11538....1C}
{Chornock} R.,  {Berger} E.,   {Fox} D.~B.,  2011, GRB Coordinates Network, \href {https://ui.adsabs.harvard.edu/abs/2011GCN.11538....1C} {11538, 1}

\bibitem[\protect\citeauthoryear{{Cline}, {Matthey}  \& {Otwinowski}}{{Cline} et~al.}{1999}]{Cline99}
{Cline} D.~B.,  {Matthey} C.,   {Otwinowski} S.,  1999, \mn@doi [\apj] {10.1086/308094}, \href {http://adsabs.harvard.edu/abs/1999ApJ...527..827C} {527, 827}

\bibitem[\protect\citeauthoryear{{Clowes}, {Harris}, {Raghunathan}, {Campusano}, {S{\"o}chting}  \& {Graham}}{{Clowes} et~al.}{2013}]{clo12}
{Clowes} R.~G.,  {Harris} K.~A.,  {Raghunathan} S.,  {Campusano} L.~E.,  {S{\"o}chting} I.~K.,   {Graham} M.~J.,  2013, \mn@doi [\mnras] {10.1093/mnras/sts497}, \href {http://adsabs.harvard.edu/abs/2013MNRAS.429.2910C} {429, 2910}

\bibitem[\protect\citeauthoryear{{Cucchiara}, {Chornock}, {Fox}  \& {Berger}}{{Cucchiara} et~al.}{2010}]{2010GCN.10422....1C}
{Cucchiara} A.,  {Chornock} R.,  {Fox} D.~B.,   {Berger} E.,  2010, GRB Coordinates Network, \href {https://ui.adsabs.harvard.edu/abs/2010GCN.10422....1C} {10422, 1}

\bibitem[\protect\citeauthoryear{{Fujii}}{{Fujii}}{2022a}]{2022SerAJ.204...29F}
{Fujii} H.,  2022a, \mn@doi [Serbian Astronomical Journal] {10.2298/SAJ2204029F}, \href {https://ui.adsabs.harvard.edu/abs/2022SerAJ.204...29F} {204, 29}

\bibitem[\protect\citeauthoryear{{Fujii}}{{Fujii}}{2022b}]{2022AN....34320021F}
{Fujii} H.,  2022b, \mn@doi [Astronomische Nachrichten] {10.1002/asna.20220021}, \href {https://ui.adsabs.harvard.edu/abs/2022AN....34320021F} {343, e20021}

\bibitem[\protect\citeauthoryear{{G{\'o}rski}, {Hivon}, {Banday}, {Wandelt}, {Hansen}, {Reinecke}  \& {Bartelmann}}{{G{\'o}rski} et~al.}{2005}]{2005ApJ...622..759G}
{G{\'o}rski} K.~M.,  {Hivon} E.,  {Banday} A.~J.,  {Wandelt} B.~D.,  {Hansen} F.~K.,  {Reinecke} M.,   {Bartelmann} M.,  2005, \mn@doi [\apj] {10.1086/427976}, \href {https://ui.adsabs.harvard.edu/abs/2005ApJ...622..759G} {622, 759}

\bibitem[\protect\citeauthoryear{{Gott}, {Juri{\'c}}, {Schlegel}, {Hoyle}, {Vogeley}, {Tegmark}, {Bahcall}  \& {Brinkmann}}{{Gott} et~al.}{2005}]{Gott05}
{Gott} III J.~R.,  {Juri{\'c}} M.,  {Schlegel} D.,  {Hoyle} F.,  {Vogeley} M.,  {Tegmark} M.,  {Bahcall} N.,   {Brinkmann} J.,  2005, \mn@doi [\apj] {10.1086/428890}, \href {http://adsabs.harvard.edu/abs/2005ApJ...624..463G} {624, 463}

\bibitem[\protect\citeauthoryear{{Haslbauer}, {Kroupa}  \& {Jerabkova}}{{Haslbauer} et~al.}{2023}]{2023MNRAS.524.3252H}
{Haslbauer} M.,  {Kroupa} P.,   {Jerabkova} T.,  2023, \mn@doi [\mnras] {10.1093/mnras/stad1986}, \href {https://ui.adsabs.harvard.edu/abs/2023MNRAS.524.3252H} {524, 3252}

\bibitem[\protect\citeauthoryear{{Horv{\'a}th}, {Hakkila}  \& {Bagoly}}{{Horv{\'a}th} et~al.}{2014}]{hhb14}
{Horv{\'a}th} I.,  {Hakkila} J.,   {Bagoly} Z.,  2014, \mn@doi [\aap] {10.1051/0004-6361/201323020}, \href {http://adsabs.harvard.edu/abs/2014A%26A...561L..12H} {561, L12}

\bibitem[\protect\citeauthoryear{{Horv{\'a}th}, {Bagoly}, {Hakkila}  \& {T{\'o}th}}{{Horv{\'a}th} et~al.}{2015}]{hbht15}
{Horv{\'a}th} I.,  {Bagoly} Z.,  {Hakkila} J.,   {T{\'o}th} L.~V.,  2015, \mn@doi [\aap] {10.1051/0004-6361/201424829}, \href {http://adsabs.harvard.edu/abs/2015A%26A...584A..48H} {584, A48}

\bibitem[\protect\citeauthoryear{{Horvath}, {Sz{\'e}csi}, {Hakkila}, {Szab{\'o}}, {Racz}, {T{\'o}th}, {Pinter}  \& {Bagoly}}{{Horvath} et~al.}{2020}]{2020MNRAS.498.2544H}
{Horvath} I.,  {Sz{\'e}csi} D.,  {Hakkila} J.,  {Szab{\'o}} {\'A}.,  {Racz} I.~I.,  {T{\'o}th} L.~V.,  {Pinter} S.,   {Bagoly} Z.,  2020, \mn@doi [\mnras] {10.1093/mnras/staa2460}, 498, 2544

\bibitem[\protect\citeauthoryear{{Horvath}, {Racz}, {Bagoly}, {Bal{\'a}zs}  \& {Pinter}}{{Horvath} et~al.}{2022}]{2022Univ....8..221H}
{Horvath} I.,  {Racz} I.~I.,  {Bagoly} Z.,  {Bal{\'a}zs} L.~G.,   {Pinter} S.,  2022, \mn@doi [Universe] {10.3390/universe8040221}, \href {https://ui.adsabs.harvard.edu/abs/2022Univ....8..221H} {8, 221}

\bibitem[\protect\citeauthoryear{{Klebesadel}, {Strong}  \& {Olson}}{{Klebesadel} et~al.}{1973}]{1973ApJ...182L..85K}
{Klebesadel} R.~W.,  {Strong} I.~B.,   {Olson} R.~A.,  1973, \mn@doi [\apjl] {10.1086/181225}, \href {http://adsabs.harvard.edu/abs/1973ApJ...182L..85K} {182, L85}

\bibitem[\protect\citeauthoryear{{Kumar Aluri} et~al.,}{{Kumar Aluri} et~al.}{2023}]{2023CQGra..40i4001K}
{Kumar Aluri} P.,  et~al., 2023, \mn@doi [Classical and Quantum Gravity] {10.1088/1361-6382/acbefc}, \href {https://ui.adsabs.harvard.edu/abs/2023CQGra..40i4001K} {40, 094001}

\bibitem[\protect\citeauthoryear{{Li} \& {Lin}}{{Li} \& {Lin}}{2015}]{Li2015}
{Li} M.-H.,  {Lin} H.-N.,  2015, \mn@doi [\aap] {10.1051/0004-6361/201525736}, \href {https://ui.adsabs.harvard.edu/abs/2015A&A...582A.111L} {582, A111}

\bibitem[\protect\citeauthoryear{{Litvin}, {Matveev}, {Mamedov}  \& {Orlov}}{{Litvin} et~al.}{2001}]{li01}
{Litvin} V.~F.,  {Matveev} S.~A.,  {Mamedov} S.~V.,   {Orlov} V.~V.,  2001, \mn@doi [Astronomy Letters] {10.1134/1.1381609}, \href {http://adsabs.harvard.edu/abs/2001AstL...27..416L} {27, 416}

\bibitem[\protect\citeauthoryear{{Lopez}, {Clowes}  \& {Williger}}{{Lopez} et~al.}{2022}]{2022MNRASLopez}
{Lopez} A.~M.,  {Clowes} R.~G.,   {Williger} G.~M.,  2022, \mn@doi [\mnras] {10.1093/mnras/stac2204}, \href {https://ui.adsabs.harvard.edu/abs/2022MNRAS.516.1557L} {516, 1557}

\bibitem[\protect\citeauthoryear{{Magliocchetti}, {Ghirlanda}  \& {Celotti}}{{Magliocchetti} et~al.}{2003}]{mgc03}
{Magliocchetti} M.,  {Ghirlanda} G.,   {Celotti} A.,  2003, \mn@doi [\mnras] {10.1046/j.1365-8711.2003.06657.x}, \href {http://adsabs.harvard.edu/abs/2003MNRAS.343..255M} {343, 255}

\bibitem[\protect\citeauthoryear{{Mandarakas}, {Blinov}, {Casadio}, {Pelgrims}, {Kiehlmann}, {Pavlidou}  \& {Tassis}}{{Mandarakas} et~al.}{2021}]{2021A&A...653A.123M}
{Mandarakas} N.,  {Blinov} D.,  {Casadio} C.,  {Pelgrims} V.,  {Kiehlmann} S.,  {Pavlidou} V.,   {Tassis} K.,  2021, \mn@doi [\aap] {10.1051/0004-6361/202140764}, \href {https://ui.adsabs.harvard.edu/abs/2021A&A...653A.123M} {653, A123}

\bibitem[\protect\citeauthoryear{{March{\~a}} \& {Browne}}{{March{\~a}} \& {Browne}}{2021}]{2021MNRAS.507.1361M}
{March{\~a}} M.~J.~M.,  {Browne} I.~W.~A.,  2021, \mn@doi [\mnras] {10.1093/mnras/stab1872}, \href {https://ui.adsabs.harvard.edu/abs/2021MNRAS.507.1361M} {507, 1361}

\bibitem[\protect\citeauthoryear{{Meszaros}}{{Meszaros}}{2006}]{mp06}
{Meszaros} P.,  2006, Reports on Progress in Physics, \href {http://adsabs.harvard.edu/abs/2006RPPh...69.2259M} {69, 2259}

\bibitem[\protect\citeauthoryear{{M{\'e}sz{\'a}ros}, {Bagoly}  \& {Vavrek}}{{M{\'e}sz{\'a}ros} et~al.}{2000a}]{mbv00}
{M{\'e}sz{\'a}ros} A.,  {Bagoly} Z.,   {Vavrek} R.,  2000a, \aap, \href {http://adsabs.harvard.edu/abs/2000A%26A...354....1M} {354, 1}

\bibitem[\protect\citeauthoryear{{M{\'e}sz{\'a}ros}, {Bagoly}, {Horv{\'a}th}, {Bal{\'a}zs}  \& {Vavrek}}{{M{\'e}sz{\'a}ros} et~al.}{2000b}]{mesz00}
{M{\'e}sz{\'a}ros} A.,  {Bagoly} Z.,  {Horv{\'a}th} I.,  {Bal{\'a}zs} L.~G.,   {Vavrek} R.,  2000b, \mn@doi [\apj] {10.1086/309193}, \href {http://adsabs.harvard.edu/abs/2000ApJ...539...98M} {539, 98}

\bibitem[\protect\citeauthoryear{{Migkas}, {Pacaud}, {Schellenberger}, {Erler}, {Nguyen-Dang}, {Reiprich}, {Ramos-Ceja}  \& {Lovisari}}{{Migkas} et~al.}{2021}]{2021A&A...649A.151M}
{Migkas} K.,  {Pacaud} F.,  {Schellenberger} G.,  {Erler} J.,  {Nguyen-Dang} N.~T.,  {Reiprich} T.~H.,  {Ramos-Ceja} M.~E.,   {Lovisari} L.,  2021, \mn@doi [\aap] {10.1051/0004-6361/202140296}, \href {https://ui.adsabs.harvard.edu/abs/2021A&A...649A.151M} {649, A151}

\bibitem[\protect\citeauthoryear{{Pe'er}}{{Pe'er}}{2015}]{2015AdAst2015E..22P}
{Pe'er} A.,  2015, \mn@doi [Advances in Astronomy] {10.1155/2015/907321}, \href {https://ui.adsabs.harvard.edu/abs/2015AdAst2015E..22P} {2015, 907321}

\bibitem[\protect\citeauthoryear{{Perley} \& {Cenko}}{{Perley} \& {Cenko}}{2015}]{2015GCN.17616....1P}
{Perley} D.~A.,  {Cenko} S.~B.,  2015, GRB Coordinates Network, \href {https://ui.adsabs.harvard.edu/abs/2015GCN.17616....1P} {17616, 1}

\bibitem[\protect\citeauthoryear{{Perley} et~al.,}{{Perley} et~al.}{2017}]{2017MNRAS.465L..89P}
{Perley} D.~A.,  et~al., 2017, \mn@doi [\mnras] {10.1093/mnrasl/slw221}, 465, L89

\bibitem[\protect\citeauthoryear{{Racz}, {Bal{\'a}zs}, {Bagoly}, {T{\'o}th}  \& {Horv{\'a}th}}{{Racz} et~al.}{2017}]{2017AIPC.1792f0012R}
{Racz} I.~I.,  {Bal{\'a}zs} L.~G.,  {Bagoly} Z.,  {T{\'o}th} L.~V.,   {Horv{\'a}th} I.,  2017, in 6th International Symposium on High Energy Gamma-Ray Astronomy. p. 060012, \mn@doi{10.1063/1.4968995}

\bibitem[\protect\citeauthoryear{{\v R\'\i pa} \& {Shafieloo}}{{\v R\'\i pa} \& {Shafieloo}}{2017}]{R_pa_2017}
{\v R\'\i pa} J.,  {Shafieloo} A.,  2017, \mn@doi [The Astrophysical Journal] {10.3847/1538-4357/aa9708}, 851, 15

\bibitem[\protect\citeauthoryear{{\v R\'\i pa} \& {Shafieloo}}{{\v R\'\i pa} \& {Shafieloo}}{2019}]{2019MNRAS.486.3027Ripa}
{\v R\'\i pa} J.,  {Shafieloo} A.,  2019, \mn@doi [\mnras] {10.1093/mnras/stz921}, \href {https://ui.adsabs.harvard.edu/abs/2019MNRAS.486.3027R} {486, 3027}

\bibitem[\protect\citeauthoryear{{Schulze} et~al.,}{{Schulze} et~al.}{2015}]{2015ApJ...808...73S}
{Schulze} S.,  et~al., 2015, \mn@doi [\apj] {10.1088/0004-637X/808/1/73}, 808, 73

\bibitem[\protect\citeauthoryear{{Tarnopolski}}{{Tarnopolski}}{2015}]{tarno15AA}
{Tarnopolski} M.,  2015, \mn@doi [\aap] {10.1051/0004-6361/201526415}, \href {http://adsabs.harvard.edu/abs/2015A%26A...581A..29T} {581, A29}

\bibitem[\protect\citeauthoryear{{Tarnopolski}}{{Tarnopolski}}{2016}]{tarno16MNRAS}
{Tarnopolski} M.,  2016, \mn@doi [\mnras] {10.1093/mnras/stw429}, \href {http://adsabs.harvard.edu/abs/2016MNRAS.458.2024T} {458, 2024}

\bibitem[\protect\citeauthoryear{{Tarnopolski}}{{Tarnopolski}}{2019}]{2019ApJTarn}
{Tarnopolski} M.,  2019, \mn@doi [\apj] {10.3847/1538-4357/aaf1c5}, \href {https://ui.adsabs.harvard.edu/abs/2019ApJ...870..105T} {870, 105}

\bibitem[\protect\citeauthoryear{{Tarnopolski}}{{Tarnopolski}}{2022}]{2022AATarn}
{Tarnopolski} M.,  2022, \mn@doi [\aap] {10.1051/0004-6361/202038645}, \href {https://ui.adsabs.harvard.edu/abs/2022A&A...657A..13T} {657, A13}

\bibitem[\protect\citeauthoryear{Tarnopolski \& Marchenko}{Tarnopolski \& Marchenko}{2021}]{Tarnopolski_2021}
Tarnopolski M.,  Marchenko V.,  2021, \mn@doi [The Astrophysical Journal] {10.3847/1538-4357/abe5b1}, 911, 20

\bibitem[\protect\citeauthoryear{{Tugay} \& {Tarnopolski}}{{Tugay} \& {Tarnopolski}}{2023}]{2023ApJ...952....3T}
{Tugay} A.,  {Tarnopolski} M.,  2023, \mn@doi [\apj] {10.3847/1538-4357/acd9a4}, \href {https://ui.adsabs.harvard.edu/abs/2023ApJ...952....3T} {952, 3}

\bibitem[\protect\citeauthoryear{{Vavrek}, {Bal{\'a}zs}, {M{\'e}sz{\'a}ros}, {Horv{\'a}th}  \& {Bagoly}}{{Vavrek} et~al.}{2008}]{vbh08}
{Vavrek} R.,  {Bal{\'a}zs} L.~G.,  {M{\'e}sz{\'a}ros} A.,  {Horv{\'a}th} I.,   {Bagoly} Z.,  2008, \mn@doi [\mnras] {10.1111/j.1365-2966.2008.13635.x}, \href {http://adsabs.harvard.edu/abs/2008MNRAS.391.1741V} {391, 1741}

\bibitem[\protect\citeauthoryear{{Woosley}}{{Woosley}}{1993}]{woo93}
{Woosley} S.~E.,  1993, \mn@doi [\apj] {10.1086/172359}, \href {http://adsabs.harvard.edu/abs/1993ApJ...405..273W} {405, 273}

\bibitem[\protect\citeauthoryear{{Woosley} \& {Bloom}}{{Woosley} \& {Bloom}}{2006}]{wb06}
{Woosley} S.~E.,  {Bloom} J.~S.,  2006, \mn@doi [\araa] {10.1146/annurev.astro.43.072103.150558}, \href {http://adsabs.harvard.edu/abs/2006ARA%26A..44..507W} {44, 507}

\bibitem[\protect\citeauthoryear{Xue, Zhang  \& Zhu}{Xue et~al.}{2019}]{Xue_2019}
Xue L.,  Zhang F.-W.,   Zhu S.-Y.,  2019, \mn@doi [The Astrophysical Journal] {10.3847/1538-4357/ab16f3}, 876, 77

\bibitem[\protect\citeauthoryear{Zhang}{Zhang}{2018}]{zhang_2018}
Zhang B.,  2018, The Physics of Gamma-Ray Bursts.
Cambridge University Press, \mn@doi{10.1017/9781139226530}

\makeatother
\end{thebibliography}








\appendix

\section{Tables}

\begin{table*}
    \centering
    \setlength{\tabcolsep}{4pt}
    \begin{tabular}{c|c|c|c|c|c|c|c|c|c|c|c|c|c|c|c|c|c} 
    \backslashbox{n}{A}
    & 0.0628 & 0.1256 & 0.188 & 0.251 & 0.314 & 0.376 & 0.439 & 0.502 & 0.565 & 0.628 & 0.691 & 0.754 & 0.816 & 0.879 & 0.942 & 1.005 & 1.068 
 \\  \hline
3&2&2&2&3&3&3&3&3&3&3&3&3&3&3&3&3&3 \\ 
4&3&3&3&3&3&4&4&4&4&4&4&4&4&4&4&4&4 \\ 
5&4&4&4&4&4&4&4&4&4&5&5&5&5&5&5&5&5 \\ 
6&4&4&4&4&4&4&4&4&5&5&5&5&5&5&5&5&5 \\ 
7&4&4&4&4&4&5&5&5&5&5&5&6&6&6&6&6&6 \\ 
8&4&4&4&4&5&5&5&5&6&6&6&6&6&6&6&6&7 \\ 
9&4&4&4&4&5&5&5&6&6&6&6&6&7&7&7&7&7 \\ 
10&4&4&4&4&5&6&6&6&6&7&7&7&7&7&7&7&8 \\ 
11&4&5&5&5&5&6&6&7&7&7&7&8&8&8&8&8&8 \\ 
12&4&5&5&5&5&6&6&7&7&7&7&8&8&8&8&8&9 \\ 
13&4&5&5&5&6&6&6&7&7&7&8&8&8&9&9&9&9 \\ 
14&4&5&5&5&6&6&6&7&7&8&8&9&9&9&9&9&10 \\ 
15&4&5&6&6&6&7&7&8&8&8&9&9&10&10&10&10&10 \\ 
16&4&5&6&6&6&7&7&8&8&9&9&10&10&10&10&10&11 \\ 
17&4&5&6&6&6&7&7&8&8&9&9&10&10&10&10&10&11 \\ 
18&4&5&6&6&7&8&8&9&9&9&10&10&11&11&11&11&12 \\ 
19&4&5&6&6&7&8&8&9&9&10&10&11&11&11&11&12&12 \\ 
20&4&5&6&6&7&8&8&9&9&10&10&11&12&12&12&12&13 \\ 
21&4&5&7&7&7&8&8&9&9&10&10&11&12&12&12&12&13 \\ 
22&4&5&7&7&8&8&9&10&10&11&11&12&12&13&13&13&13 \\ 
23&4&5&7&7&8&8&9&10&10&11&11&12&13&13&13&13&14 \\ 
24&4&5&7&7&8&9&9&10&10&11&11&12&13&13&13&14&14
     \\
     25&4&5&7&7&8&9&9&10&10&11&11&12&13&13&14&15&15
    \end{tabular}
    \caption{This table shows the number K depending on n and the area of the circle (A, in steradian) for the northern hemisphere. See Sect.~\ref{sec:nhs} for more details. See the complete table as a supplement material.}
    \label{tab:n1}
\end{table*}

\begin{table*}
    \centering
    \setlength{\tabcolsep}{4pt}
    \begin{tabular}{c|c|c|c|c|c|c|c|c|c|c|c|c|c|c|c|c|c|c} 
    \backslashbox{n}{A}
    & 0.0628 & 0.094 & 0.1256 & 0.157 & 0.188 & 0.22 & 0.251 & 0.283 & 0.314 & 0.346 & 0.377 & 0.408 & 0.439 & 0.471 & 0.502 & 0.534 & 0.565 
 \\  \hline
3 & 1 & 1 & 1 & 1 & 1 & 0.974 & 0.99 & 0.996 & 0.998 & 1 & 1 & 1 & 1 & 1 & 1 & 1 & 1 \\
4 & 0.608 & 0.867 & 0.978 & 0.995 & 0.999 & 1 & 1 & 1 & 1 & 1 & 0.763 & 0.838 & 0.89 & 0.935 & 0.957 & 0.971 & 0.985 \\
5 & \textcolor{red}{0.021} & 0.09 & 0.197 & 0.347 & 0.489 & 0.629 & 0.772 & 0.871 & 0.94 & 0.97 & 0.991 & 0.997 & 1 & 1 & 1 & 1 & 1 \\
6 & \textcolor{blue}{0.062} & 0.205 & 0.399 & 0.622 & 0.798 & 0.907 & 0.969 & 0.989 & 0.995 & 0.999 & 1 & 1 & 1 & 1 & 1 & 1 & 0.947 \\
7 & 0.122 & 0.347 & 0.634 & 0.838 & 0.946 & 0.984 & 1 & 1 & 1 & 1 & 0.855 & 0.919 & 0.959 & 0.981 & 0.989 & 0.995 & 0.999 \\
8 & 0.2 & 0.503 & 0.8 & 0.947 & 0.989 & 0.999 & 1 & 1 & 0.862 & 0.932 & 0.98 & 0.989 & 0.998 & 1 & 1 & 0.741 & 0.82 \\
9 & 0.296 & 0.635 & 0.906 & 0.986 & 0.999 & 1 & 1 & 1 & 0.963 & 0.993 & 0.997 & 0.999 & 1 & 1 & 0.874 & 0.931 & 0.966 \\
10 & 0.407 & 0.773 & 0.961 & 0.998 & 1 & 1 & 1 & 0.983 & 0.997 & 0.598 & 0.727 & 0.832 & 0.908 & 0.968 & 0.985 & 0.994 & 0.996 \\
11 & 0.497 & 0.116 & 0.277 & 0.53 & 0.75 & 0.894 & 0.979 & 0.999 & 1 & 0.774 & 0.886 & 0.944 & 0.972 & 0.533 & 0.636 & 0.728 & 0.82 \\
12 & 0.617 & 0.17 & 0.398 & 0.683 & 0.871 & 0.964 & 0.996 & 1 & 1 & 0.906 & 0.965 & 0.989 & 0.997 & 0.745 & 0.833 & 0.893 & 0.948 \\
13 & 0.714 & 0.239 & 0.511 & 0.796 & 0.95 & 0.99 & 1 & 0.787 & 0.917 & 0.974 & 0.997 & 1 & 1 & 0.886 & 0.948 & 0.978 & 0.988 \\
14 & 0.781 & 0.319 & 0.634 & 0.877 & 0.977 & 0.999 & 1 & 0.878 & 0.97 & 0.994 & 0.999 & 1 & 1 & 0.954 & 0.984 & 0.994 & 0.997 \\
15 & 0.843 & 0.401 & 0.74 & 0.938 & 0.481 & 0.686 & 0.842 & 0.944 & 0.989 & 0.698 & 0.834 & 0.931 & 0.978 & 0.608 & 0.721 & 0.832 & 0.91 \\
16 & 0.901 & 0.476 & 0.817 & 0.971 & 0.598 & 0.795 & 0.932 & 0.984 & 0.996 & 0.82 & 0.929 & 0.98 & 0.995 & 0.763 & 0.87 & 0.936 & 0.969 \\
17 & 0.937 & 0.566 & 0.888 & 0.993 & 0.707 & 0.877 & 0.973 & 0.999 & 1 & 0.903 & 0.976 & 0.998 & 0.999 & 0.88 & 0.95 & 0.982 & 0.992 \\
18 & 0.959 & 0.639 & 0.927 & 0.998 & 0.799 & 0.932 & 0.994 & 0.745 & 0.887 & 0.967 & 0.606 & 0.751 & 0.881 & 0.458 & 0.585 & 0.708 & 0.825 \\
19 & 0.978 & 0.715 & 0.967 & 0.999 & 0.874 & 0.968 & 0.999 & 0.837 & 0.94 & 0.988 & 0.735 & 0.864 & 0.951 & 0.609 & 0.732 & 0.842 & 0.928 \\
20 & 0.987 & 0.778 & 0.979 & 0.999 & 0.919 & 0.984 & 0.999 & 0.907 & 0.975 & 0.996 & 0.851 & 0.944 & 0.99 & 0.733 & 0.85 & 0.933 & 0.974 \\
21 & 0.993 & 0.833 & 0.987 & 0.999 & 0.384 & 0.641 & 0.861 & 0.963 & 0.995 & 0.79 & 0.933 & 0.988 & 0.998 & 0.844 & 0.928 & 0.973 & 0.988 \\
22 & 0.996 & 0.877 & 0.994 & 0.999 & 0.461 & 0.732 & 0.921 & 0.984 & 0.998 & 0.885 & 0.973 & 0.996 & 0.828 & 0.447 & 0.577 & 0.711 & 0.83 \\
 
    \end{tabular}
    \caption{This table shows the bootstrap probabilities (frequencies) for a certain n and the area of the circle (A, in steradian) for the northern hemisphere. See Sect.~\ref{sec:nhs} for more details. Here we are using the same color code as Fig. \ref{fig:EszakProb} used. See the complete table as a supplement material.}
    \label{tab:nprob}
\end{table*}

\begin{table*}
    \centering
    \setlength{\tabcolsep}{4pt}
    \begin{tabular}{c|c|c|c|c|c|c|c|c|c|c|c|c|c|c|c|c|c|c} 
    \backslashbox{n}{A}
     & 0.0628 & 0.1256 & 0.188 & 0.251 & 0.314 & 0.376 & 0.439 & 0.502 & 0.565 & 0.628 & 0.691 & 0.754 & 0.816 & 0.879 & 0.942 & 1.005 & 1.068 
 \\  \hline
3 & 2 & 3 & 3 & 3 & 3 & 3 & 3 & 3 & 3 & 3 & 3 & 3 & 3 & 3 & 3 & 3 & 3 \\
4 & 3 & 3 & 3 & 4 & 4 & 4 & 4 & 4 & 4 & 4 & 4 & 4 & 4 & 4 & 4 & 4 & 4 \\
5 & 3 & 3 & 4 & 4 & 5 & 5 & 5 & 5 & 5 & 5 & 5 & 5 & 5 & 5 & 5 & 5 & 5 \\ 
6 & 3 & 3 & 4 & 4 & 5 & 5 & 5 & 5 & 5 & 5 & 5 & 5 & 5 & 5 & 5 & 5 & 5 \\
7 & 3 & 4 & 4 & 4 & 5 & 5 & 5 & 5 & 6 & 6 & 6 & 6 & 6 & 6 & 6 & 6 & 6 \\
8 & 3 & 4 & 4 & 4 & 5 & 6 & 6 & 6 & 6 & 6 & 6 & 6 & 6 & 6 & 6 & 6 & 6 \\
9 & 3 & 4 & 5 & 5 & 5 & 6 & 6 & 6 & 6 & 6 & 7 & 7 & 7 & 7 & 7 & 7 & 7 \\
10 & 3 & 4 & 5 & 5 & 5 & 6 & 6 & 6 & 6 & 6 & 7 & 7 & 7 & 7 & 7 & 7 & 7 \\
11 & 3 & 4 & 5 & 5 & 5 & 6 & 6 & 7 & 7 & 7 & 7 & 7 & 7 & 8 & 8 & 8 & 8 \\
12 & 4 & 4 & 5 & 6 & 6 & 6 & 7 & 7 & 7 & 7 & 8 & 8 & 8 & 8 & 8 & 9 & 9 \\
13 & 4 & 4 & 6 & 6 & 6 & 6 & 7 & 7 & 7 & 7 & 8 & 8 & 8 & 9 & 9 & 9 & 9 \\
14 & 4 & 4 & 6 & 6 & 6 & 7 & 8 & 8 & 8 & 8 & 9 & 9 & 9 & 9 & 9 & 10 & 10 \\
15 & 4 & 5 & 6 & 6 & 6 & 7 & 8 & 8 & 8 & 8 & 9 & 9 & 9 & 9 & 9 & 10 & 10 \\
16 & 4 & 5 & 6 & 6 & 6 & 8 & 9 & 9 & 9 & 9 & 10 & 10 & 10 & 10 & 10 & 10 & 10 \\
17 & 4 & 5 & 6 & 6 & 7 & 8 & 9 & 9 & 9 & 9 & 10 & 10 & 10 & 10 & 11 & 11 & 11 \\
18 & 5 & 5 & 6 & 6 & 7 & 8 & 9 & 9 & 9 & 9 & 10 & 10 & 10 & 10 & 11 & 11 & 11 \\
19 & 5 & 5 & 6 & 7 & 7 & 9 & 10 & 10 & 10 & 10 & 11 & 11 & 11 & 11 & 11 & 12 & 12 \\
20 & 5 & 5 & 6 & 7 & 7 & 9 & 10 & 10 & 10 & 10 & 11 & 11 & 11 & 11 & 12 & 12 & 12 \\
21 & 5 & 5 & 6 & 7 & 7 & 9 & 10 & 10 & 10 & 10 & 11 & 11 & 11 & 11 & 12 & 12 & 12 \\
22 & 5 & 6 & 7 & 8 & 8 & 9 & 10 & 10 & 10 & 10 & 11 & 11 & 12 & 12 & 12 & 13 & 13 \\
23 & 5 & 6 & 7 & 8 & 8 & 9 & 10 & 10 & 10 & 10 & 11 & 11 & 12 & 12 & 12 & 13 & 13 \\
24 & 5 & 6 & 7 & 8 & 8 & 9 & 10 & 11 & 11 & 11 & 12 & 12 & 12 & 13 & 13 & 13 & 14 \\
25 & 5 & 6 & 7 & 8 & 9 & 10 & 11 & 11 & 11 & 11 & 12 & 12 & 13 & 13 & 14 & 14 & 15 \\
26 & 6 & 7 & 8 & 9 & 9 & 10 & 11 & 11 & 11 & 11 & 12 & 13 & 13 & 13 & 14 & 14 & 15 \\
27 & 6 & 7 & 8 & 9 & 10 & 10 & 11 & 12 & 12 & 12 & 13 & 13 & 14 & 14 & 15 & 15 & 16 \\
28 & 6 & 7 & 8 & 9 & 10 & 10 & 11 & 12 & 12 & 12 & 13 & 13 & 14 & 14 & 15 & 15 & 16 \\
29 & 6 & 7 & 8 & 9 & 10 & 10 & 11 & 12 & 12 & 12 & 13 & 13 & 14 & 14 & 15 & 15 & 16 \\

    \end{tabular}
    \caption{This table shows the number K depending on n and the area of the circle (A, in steradian) for the southern hemisphere. See Sect.~\ref{sec:shs} for more details. See the complete table as a supplement material.}
    \label{tab:s1}
\end{table*}

\begin{table*}
    \centering
    \setlength{\tabcolsep}{4pt}
    \begin{tabular}{c|c|c|c|c|c|c|c|c|c|c|c|c|c|c|c|c|c|c} 
    \backslashbox{n}{A}
    & 0.0628 & 0.094 & 0.1256 & 0.157 & 0.188 & 0.22 & 0.251 & 0.283 & 0.314 & 0.346 & 0.377 & 0.408 & 0.439 & 0.471 & 0.502 & 0.534 & 0.565
 \\  \hline
 3 & 1 & 1 & 0.691 & 0.824 & 0.912 & 0.957 & 0.982 & 0.996 & 0.999 & 1 & 1 & 1 & 1 & 1 & 1 & 1 & 1 \\ 
4 & 0.587 & 0.833 & 0.951 & 0.987 & 0.997 & 0.215 & 0.291 & 0.372 & 0.476 & 0.566 & 0.648 & 0.723 & 0.78 & 0.849 & 0.891 & 0.929 & 0.952 \\ 
5 & 0.801 & 0.966 & 0.999 & 1 & 0.408 & 0.583 & 0.712 & 0.805 & 0.061 & 0.092 & 0.119 & 0.145 & 0.178 & 0.233 & 0.287 & 0.349 & 0.399 \\ 
6 & 0.937 & 0.997 & 1 & 1 & 0.691 & 0.835 & 0.923 & 0.965 & 0.211 & 0.298 & 0.378 & 0.453 & 0.533 & 0.635 & 0.716 & 0.784 & 0.836 \\ 
7 & 0.986 & 0.328 & 0.53 & 0.728 & 0.867 & 0.952 & 0.987 & 0.337 & 0.459 & 0.59 & 0.702 & 0.807 & 0.873 & 0.919 & 0.953 & 0.232 & 0.281 \\ 
8 & 0.996 & 0.456 & 0.7 & 0.88 & 0.964 & 0.992 & 0.999 & 0.554 & 0.065 & 0.099 & 0.154 & 0.206 & 0.269 & 0.361 & 0.443 & 0.515 & 0.609 \\ 
9 & 0.999 & 0.594 & 0.838 & 0.968 & 0.297 & 0.459 & 0.625 & 0.777 & 0.163 & 0.238 & 0.323 & 0.416 & 0.521 & 0.627 & 0.729 & 0.804 & 0.87 \\ 
10 & 0.999 & 0.736 & 0.925 & 0.99 & 0.446 & 0.657 & 0.82 & 0.923 & 0.289 & 0.403 & 0.532 & 0.638 & 0.742 & 0.824 & 0.908 & 0.948 & 0.968 \\ 
11 & 1 & 0.842 & 0.973 & 0.997 & 0.607 & 0.817 & 0.924 & 0.98 & 0.446 & 0.587 & 0.729 & 0.823 & 0.897 & 0.337 & 0.423 & 0.509 & 0.593 \\ 
12 & 0.52 & 0.908 & 0.99 & 0.523 & 0.75 & 0.195 & 0.305 & 0.444 & 0.618 & 0.771 & 0.872 & 0.928 & 0.388 & 0.497 & 0.595 & 0.715 & 0.794 \\ 
13 & 0.598 & 0.943 & 0.997 & 0.639 & 0.144 & 0.288 & 0.437 & 0.62 & 0.773 & 0.89 & 0.95 & 0.981 & 0.576 & 0.69 & 0.783 & 0.874 & 0.931 \\ 
14 & 0.683 & 0.964 & 0.998 & 0.746 & 0.216 & 0.391 & 0.552 & 0.745 & 0.865 & 0.958 & 0.475 & 0.607 & 0.155 & 0.226 & 0.308 & 0.404 & 0.496 \\ 
15 & 0.765 & 0.294 & 0.57 & 0.842 & 0.298 & 0.509 & 0.699 & 0.864 & 0.944 & 0.478 & 0.628 & 0.751 & 0.262 & 0.372 & 0.468 & 0.584 & 0.685 \\ 
16 & 0.83 & 0.397 & 0.692 & 0.911 & 0.421 & 0.634 & 0.815 & 0.94 & 0.977 & 0.629 & 0.758 & 0.264 & 0.052 & 0.088 & 0.132 & 0.189 & 0.262 \\ 
17 & 0.867 & 0.468 & 0.765 & 0.95 & 0.518 & 0.731 & 0.888 & 0.382 & 0.563 & 0.734 & 0.272 & 0.385 & 0.098 & 0.156 & 0.227 & 0.312 & 0.398 \\ 
18 & 0.142 & 0.554 & 0.841 & 0.971 & 0.614 & 0.833 & 0.949 & 0.501 & 0.687 & 0.848 & 0.373 & 0.513 & 0.156 & 0.244 & 0.337 & 0.443 & 0.538 \\ 
19 & 0.176 & 0.631 & 0.895 & 0.984 & 0.714 & 0.272 & 0.42 & 0.621 & 0.792 & 0.923 & 0.477 & 0.146 & \textcolor{blue}{0.037} & 0.064 & 0.101 & 0.147 & 0.205 \\ 
20 & 0.215 & 0.695 & 0.938 & 0.993 & 0.794 & 0.345 & 0.52 & 0.736 & 0.886 & 0.965 & 0.592 & 0.206 & 0.061 & 0.101 & 0.158 & 0.217 & 0.285 \\ 
21 & 0.263 & 0.758 & 0.315 & 0.621 & 0.851 & 0.427 & 0.612 & 0.819 & 0.935 & 0.984 & 0.699 & 0.284 & 0.092 & 0.153 & 0.234 & 0.32 & 0.405 \\ 
22 & 0.322 & 0.141 & 0.396 & 0.709 & 0.304 & 0.073 & 0.16 & 0.297 & 0.465 & 0.658 & 0.799 & 0.379 & 0.125 & 0.207 & 0.303 & 0.405 & 0.511 \\ 
23 & 0.36 & 0.17 & 0.464 & 0.771 & 0.355 & 0.11 & 0.221 & 0.379 & 0.573 & 0.766 & 0.874 & 0.484 & 0.189 & 0.284 & 0.397 & 0.532 & 0.646 \\ 
24 & 0.414 & 0.198 & 0.522 & 0.833 & 0.418 & 0.141 & 0.269 & 0.454 & 0.68 & 0.844 & 0.932 & 0.588 & 0.259 & 0.365 & 0.503 & 0.213 & 0.294 \\ 
25 & 0.461 & 0.26 & 0.598 & 0.871 & 0.491 & 0.192 & 0.361 & 0.562 & 0.229 & 0.384 & 0.542 & 0.217 & 0.066 & 0.122 & 0.191 & 0.28 & 0.381 \\ 
26 & \textcolor{blue}{0.046} & \textcolor{blue}{0.035} & 0.12 & 0.312 & 0.108 & \textcolor{blue}{0.031} & 0.071 & 0.157 & 0.303 & 0.48 & 0.635 & 0.288 & 0.105 & 0.169 & 0.254 & 0.368 & 0.484 \\ 
27 & 0.054 & \textcolor{blue}{0.048} & 0.156 & 0.366 & 0.134 & \textcolor{blue}{0.045} & 0.104 & 0.206 & 0.058 & 0.135 & 0.24 & 0.36 & 0.147 & 0.221 & 0.332 & 0.124 & 0.183 \\ 
28 & 0.063 & 0.055 & 0.192 & 0.427 & 0.169 & 0.062 & 0.139 & 0.253 & 0.086 & 0.172 & 0.305 & 0.446 & 0.19 & 0.287 & 0.417 & 0.181 & 0.254 \\

    \end{tabular}
    \caption{This table shows the bootstrap probabilities (frequencies) for a certain n and the area of the circle (in steradian) for the southern hemisphere. 
    Here we are using the same color code as Fig. \ref{fig:delprob} used.
    See Sect.~\ref{sec:shs} for more details. See the complete table as a supplement material.}
    \label{tab:sprob}
\end{table*}

\bsp	
\label{lastpage}
\end{document}